\documentclass[11pt]{article}
\pdfoutput=1

\usepackage{jheppub}
\usepackage{url,comment}
\usepackage{times}
\usepackage{latexsym}
\usepackage{graphicx, graphics, hyperref, amsmath, amssymb, slashed, xcolor, bbm,bm,amsthm, array}
 \usepackage{subfigure}
 \usepackage{listings} 
 \usepackage{trimclip}

\usepackage{rotating}
\usepackage{afterpage}
 
\newcommand{\nc}{\newcommand}

\nc{\beq}{\begin{equation}}
\nc{\eeq}{\end{equation}}
\nc{\barray}{\begin{eqnarray}}
\nc{\earray}{\end{eqnarray}}
\nc{\barrayn}{\begin{eqnarray*}}
\nc{\earrayn}{\end{eqnarray*}}
\nc{\bcenter}{\begin{center}}
\nc{\ecenter}{\end{center}}
\nc{\mc}{\mathcal}
\nc{\er}[1]{(\ref{eq:#1})}
\nc{\onehalf}{\frac{1}{2}} 
\nc{\partialbar}{\bar{\partial}}
\nc{\psit}{\widetilde{\psi}}
\nc{\Tr}{\mbox{Tr}}
\nc{\hc}{\mbox{H.c.}}
\nc{\ev}{\;\mathrm{eV}}
\nc{\mev}{\;\mathrm{MeV}}
\nc{\gev}{\;\mathrm{GeV}}
\nc{\kev}{\;\mathrm{keV}}
\nc{\tev}{\;\mathrm{TeV}}

\def\chii0{\chi_i^0}
\def\chij0{\chi_j^0}

\newcommand{\gsim}{\lower.7ex\hbox{$\;\stackrel{\textstyle>}{\sim}\;$}}
\newcommand{\lsim}{\lower.7ex\hbox{$\;\stackrel{\textstyle<}{\sim}\;$}}
\nc{\ttbar}{t\bar t}

\newcommand{\fref}[1]{Fig.~\ref{f.#1}}
\newcommand{\eref}[1]{Eq.~(\ref{e.#1})}

\newcommand{\sref}[1]{Section~\ref{s.#1}}

\newcommand{\cref}[1]{Chapter~\ref{c.#1}}

\graphicspath{{plots/}}

\title{Cosmological Signatures of a Mirror Twin Higgs}

\author[a,b]{Zackaria Chacko,}
\author[c]{David Curtin,}
\author[a]{Michael Geller,}
\author[a]{Yuhsin Tsai}

\affiliation[a]{Maryland Center for Fundamental Physics, Department of Physics,
University of Maryland, College Park, MD 20742-4111 USA}
\affiliation[b]{Theoretical Physics Department, Fermilab, P.O. Box 500,
Batavia, IL 60510, USA}
\affiliation[c]{Department of Physics, University of Toronto, Toronto, ON M5S 1A7, Canada}

\emailAdd{zchacko@umd.edu}
\emailAdd{dcurtin@physics.utoronto.ca}
\emailAdd{mlgeller@umd.edu}
\emailAdd{yhtsai@umd.edu}

\abstract{
We explore the cosmological signatures associated with the twin 
baryons, electrons, photons and neutrinos in the Mirror Twin Higgs 
framework. We consider a scenario in which the twin baryons constitute a 
subcomponent of dark matter, and the contribution of the twin photon and 
neutrinos to dark radiation is suppressed due to late asymmetric 
reheating, but remains large enough to be detected in future cosmic 
microwave background (CMB) experiments. We show that this framework can 
lead to distinctive signals in large scale structure and in 
the cosmic microwave background. Baryon acoustic oscillations in the 
mirror sector prior to recombination lead to a suppression of structure 
on large scales, and leave a residual oscillatory pattern in the matter 
power spectrum. This pattern depends sensitively on the relative 
abundances and ionization energies of both twin hydrogen and helium, and 
is therefore characteristic of this class of models. Although both 
mirror photons and neutrinos constitute dark radiation in the early 
universe, their effects on the CMB are distinct. This is because prior 
to recombination the twin neutrinos free stream, while the twin photons 
are prevented from free streaming by scattering off twin electrons. In 
the Mirror Twin Higgs framework the relative contributions of these two 
species to the energy density in dark radiation is predicted, leading to 
testable effects in the CMB. 
These highly distinctive cosmological signatures may allow this 
class of models to be discovered, and distinguished from more general 
dark sectors.
 }

\begin{document}

\begin{flushright}
\small{.}
\end{flushright}

\maketitle

%%%%%%%%%%%%%%%%%
\section{Introduction}\label{s.introduction}
%%%%%%%%%%%%%%%%%

Mirror Twin Higgs (MTH) models 
\cite{Chacko:2005pe,Barbieri:2005ri,Chacko:2005vw} provide an 
interesting and distinctive approach to the little hierarchy problem. In 
their original incarnation, these theories contain a mirror (``twin") 
sector that has exactly the same particle content and gauge interactions 
as the Standard Model (SM). A discrete $Z_2$ twin symmetry interchanges 
the particles and interactions of the two sectors and ensures that the 
loop corrections to the Higgs mass from the mirror particles cancel the 
problematic quadratic divergences from the SM gauge and top loops. In 
contrast to most other symmetry-based solutions of the little hierarchy 
problem, the twin particles are not charged under the SM gauge groups, 
and are therefore not subject to the strong constraints from top partner 
searches at the Large Hadron Collider (LHC)~\cite{Aaboud:2017ayj, Aaboud:2017nfd, Aaboud:2017ejf, Aaboud:2016tnv, Sirunyan:2017pjw, Sirunyan:2017cwe, Khachatryan:2017rhw}.

The only coupling between the SM and the mirror sector that is required by 
the Twin Higgs mechanism is a Higgs portal interaction between the SM 
Higgs doublet and its counterpart in the the twin sector. After 
electroweak symmetry breaking, the Higgs bosons in the two sectors mix. 
This mixing leads to a suppression of the couplings of the Higgs 
particle to SM states. In addition, the Higgs can now decay into 
invisible twin sector states. Both these effects result in a reduction 
of the number of Higgs events at the Large Hadron Collider (LHC) as 
compared to the SM prediction~\cite{1992MPLA....7.2567F}. At present, this 
is the strongest collider constraint on the MTH model. In order to 
satisfy this constraint, we require a mild hierarchy between the scale 
of electroweak symmetry breaking in the twin sector, denoted by $v_B$, 
and the corresponding scale in the SM sector $v_A$, so that $v_B/v_A 
\gtrsim 3$~\cite{Burdman:2014zta}. This hierarchy can be realized by 
introducing a soft explicit breaking of the twin symmetry, albeit at the 
expense of mild tuning of order $ 2 v_A^2/ (v_A^2 + v_B^2) \simeq 20\%$. 
Then the elementary fermions and gauge bosons in the twin sector are heavier by a 
factor of $v_B/v_A$ than their SM counterparts. 

MTH models stabilize the hierarchy up to scales of order 5-10 TeV, 
beyond which an ultraviolet completion is required. Ultraviolet 
completions have been constructed based on supersymmetry 
\cite{Falkowski:2006qq,Chang:2006ra,Craig:2013fga,Katz:2016wtw,Badziak:2017syq,Badziak:2017kjk,Badziak:2017wxn} 
(for early work along the same lines, see~\cite{Berezhiani:2005ek}) and 
on the composite Higgs framework 
\cite{Geller:2014kta,Barbieri:2015lqa,Low:2015nqa} that raise the cutoff 
to the Planck scale. In supersymmetric UV completions, the breaking of 
the global symmetry is realized linearly. Then the radial mode in the 
Higgs potential is present in the spectrum, and can be searched for at 
colliders 
\cite{Buttazzo:2015bka,Katz:2016wtw,Ahmed:2017psb,Chacko:2017xpd}. In 
general, composite Twin Higgs models predict new exotic states that 
carry both SM and mirror gauge charges, which can potentially be 
discovered at the LHC~\cite{Cheng:2015buv,Cheng:2016uqk,Li:2017xyf}. 
These theories have been shown to be consistent with precision 
electroweak constraints~\cite{Contino:2017moj} and flavor 
bounds~\cite{Csaki:2015gfd}.

Cosmology places severe constraints on the MTH framework. The 
contribution of the light twin neutrinos and twin photons to the energy 
density of the universe speeds up the Hubble expansion, and this effect 
is tightly constrained by the existing CMB data. The Higgs portal 
interaction keeps the SM and the twin sector in thermal equilibrium until 
temperatures of order a GeV \cite{Barbieri:2005ri}. Below this 
temperature the twin sector continues to contribute almost half of the 
total energy density. Even after the other states in the twin sector 
have decoupled, the twin photon and neutrinos survive as thermal relics, 
resulting in a large contribution to the energy density in dark 
radiation during the CMB epoch, $\Delta N_{eff}=5.7$ 
\cite{Chacko:2016hvu,Craig:2016lyx}. A correction of this magnitude is 
ruled out by the current CMB constraints, which require $\Delta 
N_{eff}\lsim0.45\,(2\sigma)$ 
\cite{Ade:2015xua,Baumann:2015rya,Brust:2017nmv}.

Several ideas have been put forward to address this problem.{\footnote{ 
This issue can be avoided if the reheat temperature after inflation lies 
at or below a GeV, with the inflaton decaying preferentially to the visible 
sector~\cite{Ignatiev:2000yy}. However, it is not simple to explain the 
origin of the baryon asymmetry within such a framework, and we do not 
consider it further.} One approach is to admit hard breaking of the 
discrete $Z_2$ symmetry in the Yukawa couplings in the twin sector. This 
allows a large reduction in the number of degrees of freedom in the twin 
sector at the time when the two sectors decouple, leading to a 
suppression of $\Delta N_{eff}$ \cite{Farina:2015uea,Barbieri:2016zxn,Csaki:2017spo,Barbieri:2017opf}. 
More radical proposals that produce the same result involve making the 
mirror sector vector-like \cite{Craig:2016kue}, or even simply removing 
from the theory the first two generations of twin fermions, which do not 
play a role in solving the little hierarchy problem. This latter 
construction, known as the Fraternal Twin Higgs model 
\cite{Craig:2015pha}, leads to distinctive collider signatures involving 
displaced vertices that can be seen at the LHC 
\cite{Curtin:2015fna,Csaki:2015fba}, and contains several promising dark 
matter candidates 
\cite{Craig:2015xla,Garcia:2015loa,Garcia:2015toa,Freytsis:2016dgf}.

An alternative approach to resolve this problem in mirror models is to 
incorporate an asymmetric reheating process that preferentially heats up 
the SM sector rather than the mirror 
sector~\cite{Berezhiani:1995yi,Berezhiani:1995am}. In the case of MTH
models this reheating process must occur 
at late times, after the two sectors have decoupled, but before Big Bang 
nucleosynthesis (BBN). This has the effect of diluting the fraction of 
energy density contained in the twin sector, allowing the bounds on 
$\Delta N_{eff}$ to be satisfied \cite{Chacko:2016hvu,Craig:2016lyx} (see also~\cite{Adshead:2016xxj}). In 
general, late asymmetric reheating can be realized without requiring 
further breaking of the discrete $Z_2$ symmetry that relates the two 
sectors. For example, in the $\nu$MTH model \cite{Chacko:2016hvu}, 
right-handed neutrinos with mass $\mathcal{O}(10)$ GeV decouple from the 
thermal bath while still relativistic, and come to dominate the energy 
density of the universe at temperatures of order a GeV. Their decays 
occur after SM-twin decoupling, with higher branching fractions into the 
visible sector because the SM $W/Z$ bosons that mediate this process are 
lighter than their twin counterparts. This process has the effect of 
making the twin sector much colder than the SM, resulting in the 
suppression of $\Delta N_{eff}\sim 7.4(v_A/v_B)^2$. Late asymmetric reheating could also 
arise from the decay of a scalar field that is related to the 
spontaneous breaking of the ${Z}_2$ symmetry, or from $Z_2$ 
breaking in inflationary dynamics \cite{Craig:2016lyx}. Although these 
scenarios suppress $\Delta N_{eff}$, it is typically large enough to be 
seen in future CMB measurements. It is straightforward to accommodate 
baryogenesis within such a framework \cite{Farina:2016ndq}.

In this paper we explore in detail the cosmological signatures 
associated with the twin baryons, electrons, photons and neutrinos in 
the Mirror Twin Higgs scenario. We work in a framework in which the 
asymmetric twin baryon relic is assumed to constitute only a 
subcomponent of dark matter, rather than the primary component. In 
addition, the contribution of the twin photon and neutrinos to dark 
radiation is assumed to be suppressed due to late asymmetric reheating, 
but large enough to be detected in future cosmic microwave background 
(CMB) experiments. We primarily focus on the case in which the 
discrete $Z_2$ symmetry is only softly broken, so the mirror particles 
are heavier than their SM counterparts by a factor of $v_B/v_A$. Then 
the relative fractions of hydrogen and helium in the mirror sector can 
be determined as a function of $v_B/v_A$, $\Delta N_{eff}$ and the 
baryon asymmetry. We show that this class of theories gives rise to 
distinctive signals in large scale structure (LSS) and the CMB that can 
potentially be detected in future experiments.

To understand the origin of these signatures, we consider the thermal 
history of the MTH model after asymmetric reheating has occurred. Once 
the temperature of the universe falls below a few MeV, twin Big Bang 
nucleosynthesis (TBBN) begins in the mirror sector. This determines the 
relative abundances of twin hydrogen and twin helium at later times. 
Later, at temperatures below a keV, while density perturbations in cold 
dark matter (CDM) are growing logarithmically, the mirror baryons are 
scattering off the mirror electron and mirror photon, leading to twin 
baryon acoustic oscillations (TBAO). This prevents the mirror particles 
from contributing to structure growth, resulting in smaller 
inhomogeneities in the matter distribution (for modes that entered the 
horizon during the radiation dominated era) than would be expected from 
$\Lambda$CDM. At temperatures of order an eV, as the universe approaches 
the epoch of matter domination, recombination occurs in the twin sector, 
first for mirror helium and subsequently for mirror hydrogen. After this 
time, neutral mirror hydrogen and helium atoms behave as CDM and start 
to clump, contributing to structure growth. The oscillations in the 
mirror sector, apart from suppressing structure at short wavelengths, 
leave a residual oscillatory imprint in the matter power spectrum that 
can potentially be seen in future LSS measurements. In detail, this 
imprint depends sensitively on the relative abundances and ionization 
energies of both twin hydrogen and helium, and is therefore highly 
characteristic of this class of models. In the absence of a signal, 
these measurements will be able to set an upper bound on the twin baryon 
density, and the results can be translated into an upper bound on the 
twin baryon asymmetry. As we discuss later, future high precision galaxy 
redshift surveys and cosmic shear surveys \cite{Font-Ribera:2013rwa} are 
expected to be sensitive to both the suppression in the matter power 
spectrum and the oscillatory feature.

The twin neutrinos and twin photons have different effects on the CMB 
anisotropies. At early times, while the twin neutrinos free stream, the 
twin photons are prevented from free streaming by scattering off of the 
twin electrons. Consequently, the density perturbations in these two 
species evolve differently, with the result that their imprints in the 
CMB are distinct. In the MTH framework, although $\Delta N_{eff}$ itself 
is a free parameter, the relative energy densities in these two species 
are predicted. This leads to a testable prediction for the corrections 
to the heights and locations of the CMB peaks that can potentially be 
tested in future experiments.

All of these characteristic cosmological signals of the MTH framework, 
including dark BAO and contributions to $\Delta N_{eff}$, are in fact 
features of the more general class of models in which the states in a 
mirror sector constitute some or all of the observed dark matter. 
Reviews of the status of mirror dark matter, which include many 
references, may be found in 
\cite{Berezhiani:2003xm,Ciarcelluti:2010zz,Foot:2014mia}. In detail, 
however, these signals depend sensitively on the masses of the mirror 
particles and the temperature in that sector. From this perspective, our 
paper therefore represents an updated, detailed study of the 
cosmological signals of mirror models, in the region of parameter space 
motivated by the hierarchy problem.
 
In this paper our primary focus is on the case in which the discrete 
$Z_2$ symmetry is only softly broken, which leads to a prediction for 
the relative abundances of twin hydrogen and helium as a function of 
$v_B$ and the temperature of the mirror sector. However, for the 
purposes of comparison we also study the scenarios in which, for a given 
electron mass, the nuclei in the mirror sector are composed entirely of 
hydrogen, or entirely of helium. These studies therefore provide some 
insight into the cosmology of MTH models in which the Yukawa couplings 
of the light quarks exhibit hard breaking of the discrete $Z_2$ 
symmetry, so that the spectrum of mirror nuclei is composed of only a 
single species, either hydrogen or helium. In particular, this allows us 
to capture the cosmological signatures of the interesting scenario in 
which the mirror neutron is lighter than the mirror proton, and 
constitutes the primary component of the observed dark matter 
\cite{Barbieri:2005ri,Addazi:2015cua,Barbieri:2017opf}, while mirror 
helium represents an acoustic subcomponent that gives rise to the 
signals we discuss. This scenario can also arise in two Higgs doublet 
extensions of mirror models even in the absence of hard $Z_2$ breaking 
if tan$\beta$, the ratio of the VEVs of the two doublets, is different 
in the two sectors~\cite{Berezhiani:2000gh}. As a dark matter candidate, 
mirror neutrons have the attractive feature that their self-interactions 
are parametrically of the right size to explain the observed small scale 
cosmological anomalies~\cite{Spergel:1999mh}. Interestingly, we find 
that the LSS of the framework in which both hydrogen and helium are 
present exhibits distinctive features that may allow it to be 
distinguished from the case of atomic dark matter with just a single 
type of nucleus.

The twin sector may already be playing a role in resolving some existing 
cosmological puzzles~\cite{Prilepina:2016rlq}. For almost two decades, 
the $\Lambda$CDM model has provided an excellent fit to cosmological 
data on large scales. However, with the advent of precision 
measurements, the standard paradigm has come into tension with the data. 
In particular, there is a $\sim 3\sigma$ discrepancy between the value 
of the Hubble rate $H_0$ obtained from a fit to the CMB and baryon 
acoustic oscillation (BAO) data \cite{Ade:2015xua} and the results from 
local measurements \cite{Riess:2016jrr}. In addition, the inferred value 
of the parameter $\sigma_8$, which corresponds to the amplitude of 
matter density fluctuations at a scale of $8h^{-1}$ Mpc, is in 
$2$-$3\sigma$ tension with the direct measurements obtained from weak 
lensing surveys~\cite{Heymans:2013fya,Hildebrandt:2016iqg}. Although the 
recently published results from the Dark Energy Survey (DES) data 
\cite{Abbott:2017wau} exhibit smaller disagreement with the Planck 
results (less then $2\sigma$), the fact that the low redshift 
measurements of $\sigma_8$ consistently give lower values is intriguing. 
Resolving these anomalies would require a framework that generically 
reduces the value of $\sigma_8$ as compared to the $\Lambda$CDM model, 
while enhancing $H_0$. Several ideas have been proposed that make use of 
a non-minimal dark sector to address these problems
\cite{Lesgourgues:2015wza,Ko:2016uft,Chacko:2016kgg,Prilepina:2016rlq,Ko:2017uyb,Buen-Abad:2017gxg,Raveri:2017jto}, 
and the MTH appears to possess all of the necessary features.

The outline of this paper is as follows. In \sref{modelparams} we 
describe the early thermal history of the universe within the MTH 
framework. A set of model inputs that parameterizes the cosmology of the 
mirror sector is defined in \sref{inputparams}. These inputs also 
determine the mass splitting between the twin proton and neutron, which 
is important for TBBN. In Sec.~\ref{s.BBN} we compute the neutron-proton 
density ratio right before TBBN, which can be translated into the 
relative fractions of twin hydrogen and twin helium. In Sec.~\ref{s.rec} 
we discuss the physics relevant for twin recombination, and derive the 
ionization fraction of the twin electron as a function of the mirror 
baryon density. In Sec.~\ref{s.cmblss}, we study TBAO and estimate the 
suppression of the matter power spectrum resulting from oscillations in 
the mirror sector. We obtain a constraint on the twin baryon density 
using current results from LSS measurements, and discuss the future 
observation of oscillation patterns from TBAO. In Sec.~\ref{s.CMB}, we 
discuss the distinct effects of twin neutrinos and twin photons on the 
CMB spectrum, and show that this leads to a testable prediction. Our 
conclusions are in Sec.~\ref{s.conclusion}.

%%%%%%%%%%%%%%%%%
\section{Thermal History}\label{s.modelparams}
%%%%%%%%%%%%%%%%%

In this section we describe the early thermal history of the universe 
within the MTH framework. This will set the stage for the computation of 
LSS and CMB observables.

\subsection{Input Parameters}
\label{s.inputparams}
%%%%%%%

 Our focus is on the cosmological signatures of MTH models in which the 
twin baryons constitute a subcomponent of dark matter. We restrict our 
analysis to the case when the Yukawa couplings respect the discrete 
$Z_2$ symmetry that relates the two sectors, so that the twin fermions 
are heavier than their visible counterparts by a factor of $v_B/v_A$. 
The energy density in twin radiation is assumed to be diluted by late 
time asymmetric reheating after the two sectors have decoupled, allowing 
the current CMB and BBN constraints to be satisfied. For simplicity we 
neglect the masses of both the SM and twin neutrinos. In this framework, 
the effects on late time cosmology are determined by the following three 
parameters,
 \begin{equation}\label{eq:rall}
\Delta N_{eff}, \qquad v_B/v_A, 
\qquad r_\mathrm{all}  = \Omega_{\text{all\ mirror\ baryons}}/\Omega_{\text{DM}}.
 \end{equation}
 Here $\Delta N_{eff}$ represents the energy density in twin radiation 
parametrized in terms of the effective number of neutrinos, while 
$r_\mathrm{all}$ denotes the \emph{total} 
asymmetric mirror baryon density relative to the total dark matter density today. Given these three 
parameters, we can determine from TBBN the fractional contributions of 
twin hydrogen and twin helium to the total dark matter energy density, 
$r_{\hat{\text{H}}}$ and $r_{\hat{\text{H}}\text{e}}$,
 \begin{equation}
r_{\hat{\text{H}}}  = \Omega_{\hat{\text{H}}}/\Omega_{\text{DM}}, 
\qquad
r_{\hat{\text{H}}\text{e}}  = \Omega_{\hat{\text{H}}\text{e}}/\Omega_{\text{DM}}
\;.
 \end{equation} The magnitude of $\Delta N_{eff}$ depends on the details 
of the asymmetric reheating process that occurs after the two sectors 
have decoupled, and so we simply treat it as an input parameter for our 
study. However, the relative contributions of the twin photons and twin 
neutrinos to $\Delta N_{eff}$ are independent of the nature of the 
reheating process. During the CMB era, prior to recombination, while the 
twin neutrinos free stream, the twin photons scatter off the ionized 
twin electrons. Consequently, as we discuss in Sec.~\ref{s.CMB}, the 
inhomogeneities associated with these two species do not evolve in the 
same way, and so their effects on the CMB are different. Then the fact 
that the relative energy densities in these two species are known leads 
to a prediction that can potentially be tested in future CMB 
experiments.

For late time cosmology, the masses of the twin electrons and twin 
baryons are especially important. These depend on the ratio $v_B/v_A$. 
While Higgs coupling measurements at the LHC constrain $v_B/v_A \gtrsim 
3$, the requirement that the Higgs mass be only modestly tuned limits 
$v_B/v_A \lesssim 5$. The mass of the twin electron is simply $v_B/v_A$ 
times the corresponding value in the SM. To determine 
the masses of the twin baryons, note that the quark masses are also $v_B/v_A$ 
times larger than in the SM. This affects the running of the mirror 
QCD coupling, and leads to a larger confinement scale in the mirror 
sector than in the SM by 30-50\% for $v_B/v_A=3$-$5$. The proton and neutron 
masses, which are almost entirely dictated by $\Lambda_{QCD}$, scale the 
same way in the mirror sector,
 \begin{equation}
\label{e.LambdaQCDB}
\frac{
m_{\hat{p}}
}{m_{p}} 
\approx
\frac{
m_{\hat{n}}
}{m_{n}}
\approx
\frac{\Lambda_{QCD_B}}{\Lambda_{QCD_A}}
\approx 
0.68 + 0.41 \log(1.32 + v_B/v_A)
 \end{equation} 
 The function of $v_B/v_A$ in Eq.~(\ref{e.LambdaQCDB}) is a numerical fit 
valid in the range $v_B/v_A \in (2,10)$ that gives excellent agreement 
with the solution from the $\overline{\rm MS}$ 1-loop RGEs. Given the twin proton 
mass $m_{\hat{p}}$, we can relate $r_{{\text{all}}}$ to 
the baryon asymmetry in the twin sector $\eta_{\hat{b}}$, 
 \begin{equation}
\frac{\eta_{\hat{b}}}{\eta_{b}}=\frac{r_{\text{all}}
\,\Omega_{\text{DM}}\,m_p}{\Omega_b\,m_{\hat{p}}} \;.
 \end{equation} 
 In this expression, the baryon asymmetries are defined as the ratio of 
SM or twin baryon number density to the total entropy density of the 
universe. As in the SM, the contributions of the twin sector to the 
energy density in dark matter are almost entirely from mirror hydrogen 
and helium,
 \begin{equation}
r_\mathrm{all} = r_{\hat{\text{H}}} + r_{\hat{\text{\text{H}}}\text{e}}
 \end{equation}
 The LSS signals, as well as the twin baryon distribution in galaxies, depend on the relative abundances of mirror hydrogen 
and helium. Prior to recombination, the mirror ions, electrons and 
photons undergo dark acoustic oscillations, and do not contribute to the 
buildup of inhomogeneities. Only after recombination do the neutral 
mirror atoms contribute to structure growth. Since the ionization 
energies of mirror hydrogen and helium are different, recombination 
occurs at different times for these two species. Therefore the matter 
power spectrum in the MTH framework is very sensitive to the relative 
abundances of mirror hydrogen and helium.

The relative fractions of mirror hydrogen and helium in the early 
universe are determined by the dynamics of BBN in the twin sector. As in 
the SM, the result is very sensitive to the mass splitting between the 
proton and neutron, $\Delta M_{np}$. This mass difference depends on the 
quark masses and $\Lambda_{QCD}$ as,
 \begin{equation}
\label{e.DeltaMnporig}
\Delta M_{np} \approx C (m_d - m_u) - D \alpha_{EM} \Lambda_{QCD}.
 \end{equation}
We extract the $(C,\,D)$ coefficients from Fig.~3 of the lattice QCD study \cite{Borsanyi:2014jba}, which gives $(C,\,D)\approx (0.86,0.40)$. Then, using \eref{LambdaQCDB} and the fact that the twin quark masses 
are larger by a factor of $v_B/v_A$ than their SM counterparts, we can 
relate the neutron-proton mass differences in the two sectors,
 \begin{eqnarray}
\label{e.DeltaMnpfull}
\frac{\Delta M_{\hat{n}\hat{p}}}{\Delta M_{np}}
&\approx & 1.68 v_B/v_A - 0.68,\qquad \Delta M_{np}=1.29\,\text{MeV}.
 \end{eqnarray}
 For the range we are interested in, $3\leq v_B/v_A \leq 5$, the 
separation between twin proton and neutron masses is $\approx 5$-$12$ 
MeV. We are now in a position to determine the relative fractions of 
mirror hydrogen and mirror helium in the twin sector as a function 
of $\Delta N_{eff}$, $v_B/v_A$ and $r_\mathrm{all}$.

%%%%%%%%%%%%%%%%%
\subsection{BBN}\label{s.BBN}
%%%%%%%%%%%%%%%%%

As in the SM, the most abundant elements in the mirror sector are 
expected to be twin hydrogen and twin helium. Their relative abundance 
is determined by the dynamics of TBBN. Under the assumption that the 
mirror sector contains a baryon asymmetry leading to stable mirror 
baryon relics, the mirror helium fraction affects LSS formation and TBAO. 
A precise calculation of the light element abundances, even in 
the SM, is extremely involved \cite{Bernstein:1988ad,Sarkar:1995dd}. 
Fortunately, the helium fraction can be estimated using the much simpler 
calculation of neutron-proton freeze-out, without considering the 
nuclear reactions in detail. This allows us to determine the mirror 
helium fraction with remarkable precision, despite large uncertainties 
due to mirror nuclear physics. Earlier studies of BBN in mirror models 
may be found in 
\cite{Berezhiani:2000gw,Ciarcelluti:2014vta,Foot:2014mia}

\subsubsection{\it BBN in the SM}

We first review a simple estimate of the helium fraction in the SM, following closely the analytical procedure of~\cite{Mukhanov:2003xs} (see also e.g. \cite{Fradette:2017sdd} and \cite{Bernstein:1988ad,Sarkar:1995dd}). 
The first step of the calculation involves computing the neutron-proton ratio ``after freeze-out'' but before the onset of nuclear reactions and neutron decay. 
The $n\leftrightarrow p$ weak conversion rates are approximated by
following integrals over thermal distributions, 
\begin{equation}
\Gamma_{n\nu_e\to p e^-}=\frac{1+3g_a^2}{2\pi^3}G_F^2Q^5J(1;\infty),\quad\Gamma_{ne^+\to p \bar{\nu}_e}=\frac{1+3g_a^2}{2\pi^3}G_F^2Q^5J(-\infty;-\frac{m_e}{Q}),
\end{equation}
where 
\begin{equation}
J(a,b)\equiv\displaystyle{\int_a^b}\sqrt{1-\frac{(m_e/Q)^2}{q^2}}\frac{q^2(q-1)^2\,dq}{(1+e^{\frac{Q}{T_{\nu}}(q-1)})(1+e^{-\frac{Q}{T}q})}.
\end{equation}
The inverse reaction rates are derived from detailed balance,
\begin{equation}
\Gamma_{pe^-\to n\nu_e}=e^{-Q/T}\Gamma_{n\nu_e\to p e^-},\quad\Gamma_{p\bar{\nu}_e\to n e^+}=e^{-Q/T}\Gamma_{ne^+\to p \bar{\nu}_e}.
\end{equation}
 Here $g_a\simeq 1.27$ is the standard nucleon axial-vector coupling, 
$Q=m_n-m_p\simeq 1.293$ MeV, $G_F$ is the Fermi constant, and $J$ is 
evaluated numerically. Electrons are assumed to annihilate away in a 
step-function approximation at $T \approx m_e/20$, and the neutrino 
temperature is $T_{\nu}=T$ $(T_\nu = (4/11)^{1/3} T )$ before (after) 
electron annihilation. The differential equation for the neutron 
fraction $X_n\equiv n_n/(n_n+n_p)$ is
 \begin{equation}
\label{e.dXndT}
\frac{dX_n}{dT}=\frac{\Gamma_{n\nu_e\to p e^-}+\Gamma_{ne^+\to p \bar{\nu}_e}}{TH(T)}\left(X_n-(1-X_n)e^{-Q/T}\right) \ .
\end{equation} 
Solving this differential equation numerically, we find that $X_n$ reaches the freeze-out value 
\begin{equation}
X_{n}^\mathrm{FO} = 0.15
\end{equation}
around $T = T_{n}^\mathrm{FO} \approx 0.2 \mev$, in agreement with \cite{Bernstein:1988ad,Sarkar:1995dd}. 
That temperature corresponds to $t \sim 20\ \mathrm{s}$, much less than the neutron lifetime  $\tau_{n} \approx 880\ \mathrm{s}$, which is why we did not include a neutron decay term in \eref{dXndT}.

Following neutron-freeze-out one has to consider the onset of nuclear reactions, which eventually give rise to the light elemental abundances, as well as the competing process of neutron decay. The time scale of nucleosynthesis is dominated by the ``deuterium bottleneck'', since the formation of helium and other elements proceeds  via deuterium. The binding energy of deuterium is very small, $B_{D} \approx 2.2 \mev$, causing it to break apart at temperatures above $\sim 0.1 \mev$. Once the temperature has dropped below that threshold and deuterium is stable, the other elements form extremely rapidly at $t = t_\mathrm{ns} \approx 180\ \mathrm{s}$, with almost all of the remaining neutrons being used up to form helium. The final helium fraction can therefore be computed from the remaining neutron fraction,
\begin{equation}
\label{e.Xntns}
X_{n}(t_\mathrm{ns}) \approx X_{n}^\mathrm{FO} e^{-t_\mathrm{ns}/\tau_{n}} \approx 0.122 \ .
\end{equation}
The primordial helium mass fraction is then
\begin{equation}
Y_p(^4\mathrm{He}) \equiv 
\frac{\rho_{\text{He}}}{\rho_{\text{He}}+\rho_{\text{H}}} \approx  \frac{4 n_{\text{He}}}{n_{\text{H}} + 4 n_{\text{He}}} = 2 X_n (t_\mathrm{ns}) \approx 0.245 \ .
\end{equation}
Note that only about 20\% of neutrons decay before they are bound up in helium nuclei following freeze-out.

\subsubsection{\it {BBN in the Mirror Sector}}

The first step of this calculation can be easily repeated for the mirror sector to obtain $X_{\hat n}^\mathrm{FO}$, by replacing $m_e \to m_{\hat e} = (v_B/v_A) m_e$, $G_F \to \hat G_F = (v_B/v_A)^{-2} G_F$ and $Q \to \hat Q = \Delta M_{\hat n \hat p} =  \frac{\Delta M_{\hat n \hat p}}{ \Delta M_{np}} Q$ using \eref{DeltaMnpfull}. Since $H$ is dominated by the visible sector, $T$ still refers to the visible sector temperature, but in the integrated distribution functions $J$, the mirror sector temperature $\hat T$ must be used. This is related to the visible sector temperature by
\begin{equation}\label{eq:rT}
r_T \equiv
\frac{\hat T}{T} = \left( \frac{g_{\star A}}{g_{\star B}} \right)^{1/3} \left( \frac{\Delta N_\mathrm{eff}}{7.4}\right)^{1/4} \ < 1 .
\end{equation}
 The mirror sector neutrino temperature is given by the usual $\hat 
T_{\nu}=\hat T$, $(\hat T_\nu = (4/11)^{1/3} \hat T )$ before (after) 
mirror electron annihilation. The mirror neutron-proton ratio, 
helium-hydrogen number density ratio, and mirror helium mass fraction 
derived from the resulting $X_{\hat n}^\mathrm{FO}$ are shown in 
\fref{mirrornoverp} as a function of $v_B/v_A$ and $\Delta 
N_\mathrm{eff}$.

\begin{figure}
\begin{center}
\begin{tabular}{ccc}
\hspace{4mm}
\footnotesize $n_{\hat n} / n_{\hat p}$& 
\hspace{4mm}
\footnotesize $n_\mathrm{\hat He}/ n_\mathrm{\hat H}$ & 
\hspace{4mm}
\footnotesize $\hat Y_p(^4\mathrm{\hat He}) =  \rho_\mathrm{\hat He} / (\rho_\mathrm{\hat He}+ \rho_\mathrm{\hat H})$ 
\\
\includegraphics[width=5cm]{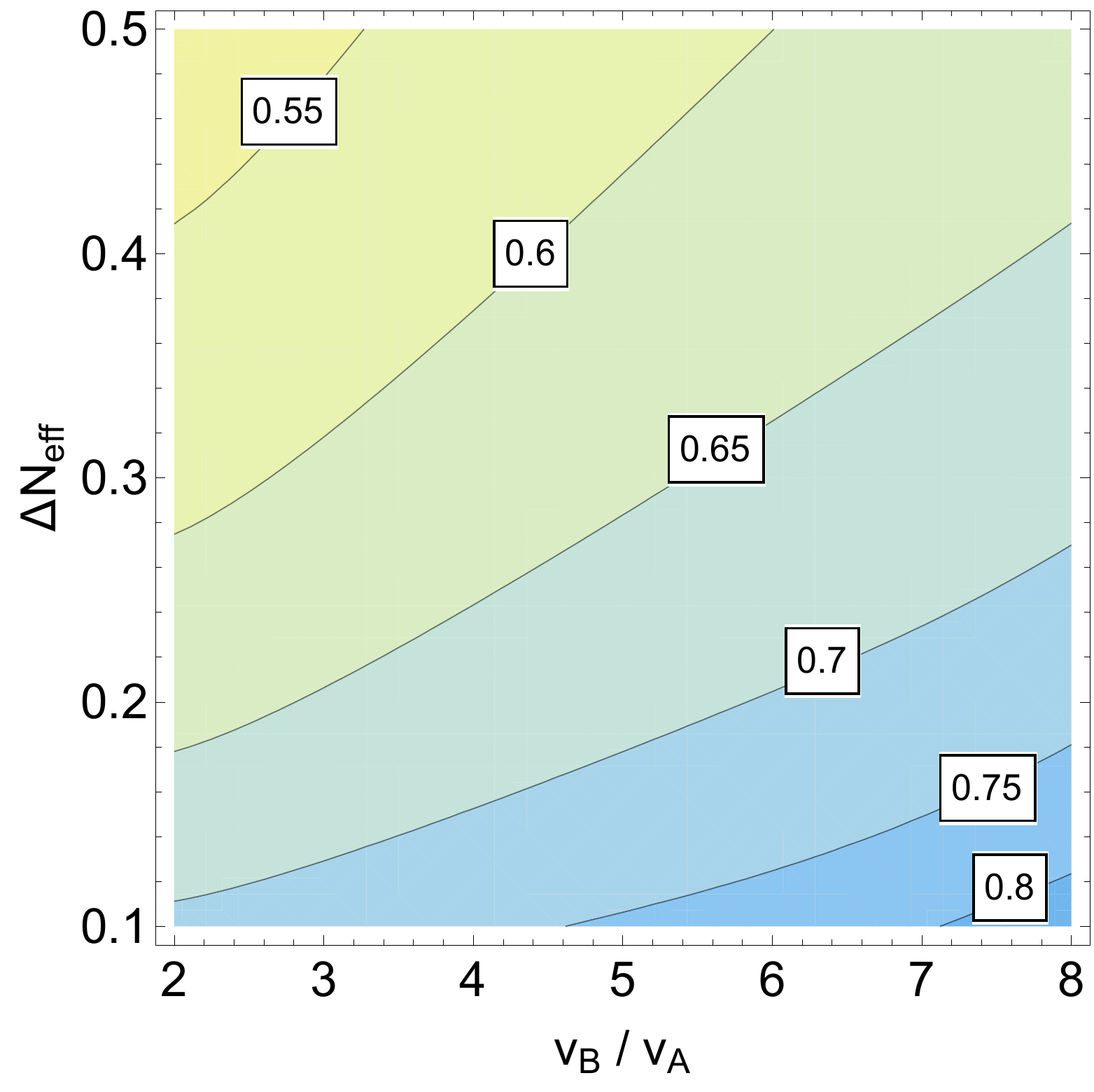} &\includegraphics[width=5cm]{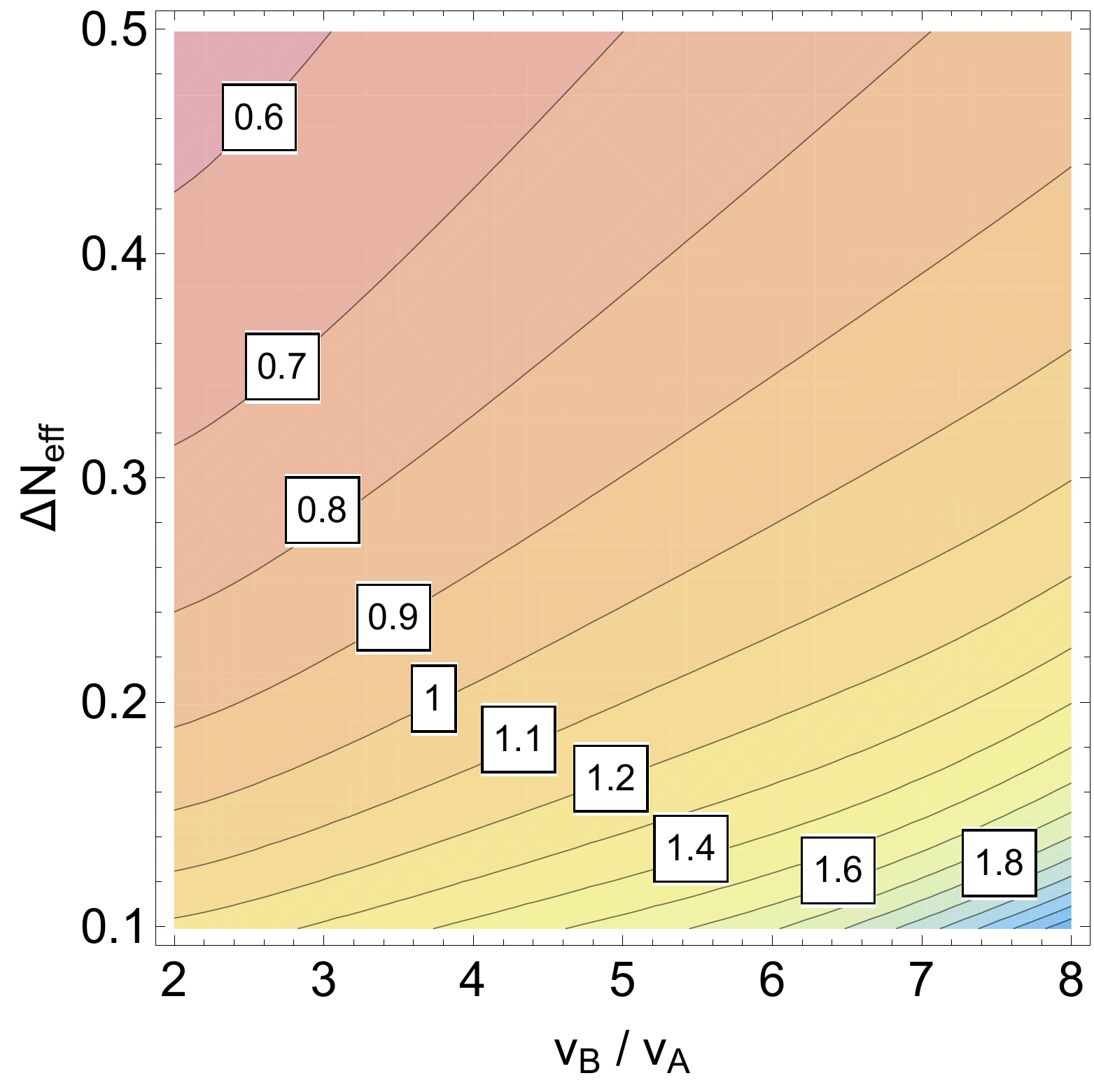} &\includegraphics[width=5cm]{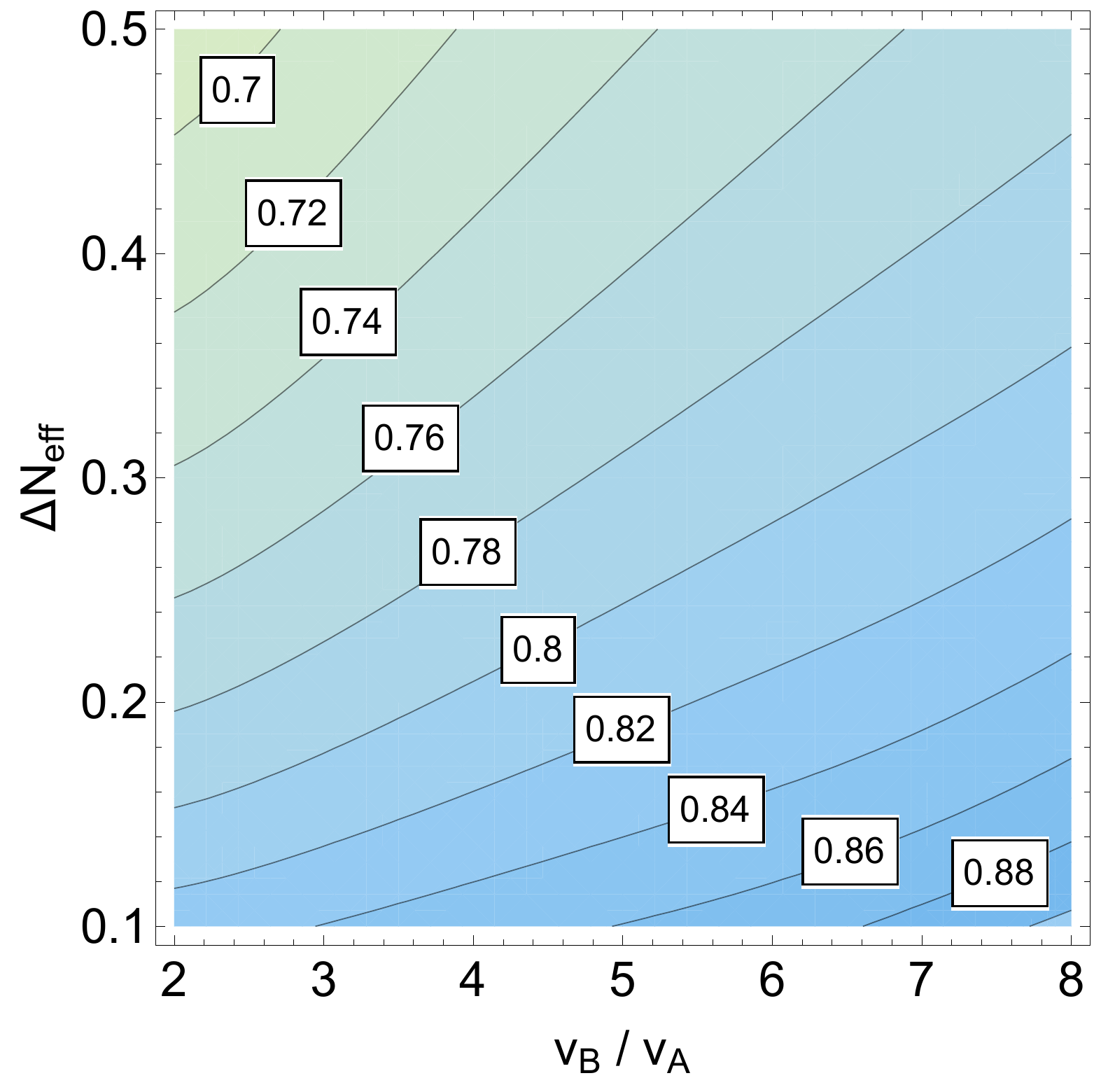}\\
\footnotesize (SM: 0.14) &\footnotesize (SM: 0.08)  &\footnotesize (SM: 0.245) 
\end{tabular}
\end{center}
\vspace*{-5mm}
\caption{From left to right: the mirror $(n/p)$ ratio, ratio of the mirror $^4$He and H number density, and ratio of the mirror $^4$He and H matter density in MTH model. We are interested in the region $3\leq v_B/v_A\leq 5$. The corresponding value in the visible sector is indicated below each plot. 
}
\label{f.mirrornoverp}
\end{figure}

Before discussing the physical ramifications of these results, we have 
to justify our use of $X_{\hat n}^\mathrm{FO}$ to directly derive the 
helium fraction. In obtaining \fref{mirrornoverp}, we assumed that 
$X_{\hat n} \approx X_{\hat n}^\mathrm{FO}$ in the mirror version of 
\eref{Xntns}. This corresponds to the assumption that mirror neutron 
decay is much slower than the onset of TBBN and can be neglected. This 
requires that the mirror deuteron binding energy $B_{\hat D}$ is 
significantly larger than in the SM, even relative to the shorter mirror 
neutron lifetime.\footnote{We also assume that the mirror baryon density 
is within a few orders of magnitude of the SM baryon density, so as not 
to prohibitively suppress the rate of nucleon collisions.} We now argue 
that this is indeed the case.

Because of its unnaturally small binding energy, the deuteron is 
understood to be a fine tuned 
system~\cite{Weinberg:1990rz,Weinberg:1991um,Bedaque:2002mn}. While this 
prevents us from calculating the mirror deuteron binding energy 
analytically, we can reuse lattice calculations of the SM deuteron 
binding energy for different pion masses to obtain an estimate, 
see~\cite{Savage:2015eya} for a review. The lattice studies find that 
the binding energy increases with pion mass, but only a handful of such 
calculations have been performed, and different methods appear to yield 
somewhat different results. Even so, we can bracket the range of 
possibilities for the binding energy as a function of pion mass with two 
linear parameterizations:
 \begin{eqnarray}
\nonumber
B_{ D}^\mathrm{min} & = & - (0.66 \mev) + 0.021\, m_{ \pi} \ , \\
\label{e.BDhat}
B_{ D}^\mathrm{max} & = & - (9.2 \mev) + 0.084 \,m_{ \pi}  \ ,
 \end{eqnarray}
both of which reduce to $B_D = 2.2 \mev$ when $m_{ \pi} = 135 \mev$. 
To apply this parameterization to the mirror sector, we rescale the dimensionful constant by $\hat \Lambda_\mathrm{QCD}/\Lambda_\mathrm{QCD}$, see \eref{LambdaQCDB}, and replace $m_\pi$ by $m_{\hat \pi}$, which is given by 
\begin{equation}
m_{\hat \pi} = \sqrt{\frac{ \hat \Lambda_\mathrm{QCD}}{\Lambda_\mathrm{QCD}} \frac{v_B}{v_A}} \ m_\pi
\approx
\sqrt{\left[0.68 + 0.41 \log(1.32 + v_B/v_A)\right] \frac{v_B}{v_A}} \ m_\pi \ .
\end{equation}
Now define $T_n$ ($T_{\hat n}$) as the \emph{visible} sector temperature when the SM (mirror) neutron decay width equals the Hubble expansion: $\Gamma_n$ ($\Gamma_{\hat n}$) $ = H(T)$. This is related by a numerical factor (same in the visible and mirror sectors) to the temperature at which neutrons would typically decay. 
Note that $\Gamma_n \propto \Delta M_{np}^5/v^4$, $\Gamma_{\hat n} = \Gamma_n  \left(\Delta M_{\hat n \hat p}/{\Delta M_{np}}\right)^5 \left(v_B/v_A\right)^{-4}$, where the ratio of $n-p$ mass splittings is given in \eref{DeltaMnpfull}.
We also define $T_D$ ($T_{\hat{D}}$) as the \emph{visible} sector temperature when the SM (mirror) deuterium bottleneck is resolved and SM (mirror) BBN starts. Up to a common prefactor, this is given by
\begin{eqnarray*}
T_D = B_D \ \ , \ \ \ 
T_{\hat D} = \frac{B_{\hat D}}{r_T} \ .
\end{eqnarray*}
Since $t \sim T^{-2}$, 
\begin{equation}
\frac{t_{ns}}{\tau_n}
 \propto
\left(\frac{T_n}{T_D}\right)^2 \ ,
\end{equation}
and hence
\begin{equation}
\frac{\hat t_{ns}}{\tau_{\hat n}} = r_{nD} \frac{t_{ns}}{\tau_n} ,\qquad r_{nD} \equiv \left(\frac{ T_{\hat n}/T_{\hat D}}{ T_n/T_D}\right)^2 \ .
\end{equation}
Recall how in the SM calculation \eref{Xntns}, $X_n^\mathrm{FO}$ is scaled down by $e^{-t_\mathrm{ns}/\tau_{n}} \approx 0.8$ to give the final neutron and hence helium yield. 
Therefore, $r_{nD}$ parameterizes how important neutron decay is for TBBN. If $r_{nD} < 1$, neutron decay is \emph{less} important in the mirror sector than in the SM. 

We can compute $r_{nD}$ for each of the two parameterizations of $B_{\hat D}$ in \eref{BDhat}. 
The dependence of $r_{nD}$ on $v_B/v_A$ and $\Delta N_\mathrm{eff}$ is very modest, much weaker than the dependence on the parameterizations of the mirror deuteron binding energy. For $B_D = B_D^\mathrm{max\ (min)}$, $r_{nD} \approx \frac{1}{16} \ (1)$. The final neutron yield $X_{\hat n}$ must then satisfy
\begin{equation}
0.8 \lesssim \frac{X_{\hat n}}{X_{\hat n}^\mathrm{FO}} \lesssim 0.8^{1/16} \approx 1 \ .
\end{equation}
Unless the mirror deuteron binding energy is very close to our minimum estimate, $X_{\hat n}/{X_{\hat n}^\mathrm{FO}}$ will be very close to 1. This justifies our use of $X_{\hat n}^\mathrm{FO}$ to estimate the helium fraction in \fref{mirrornoverp}. At worst, $n_{\hat n}/n_{\hat p}$,  $n_\mathrm{\hat He}/n_\mathrm{\hat H}$ and $\hat Y_p(^4\mathrm{\hat He})$ will be lower than the values shown in \fref{mirrornoverp} by about 20\%, 40\% and 10\% respectively, which will not significantly change our conclusions.

\fref{mirrornoverp} allows us to make a remarkable prediction for the 
MTH model. The primordial mirror neutron-to-proton ratio is $\sim 
0.6\,$-$\,0.7$, compared to the SM value of 0.14. As a result, the 
mirror helium mass fraction is $\hat Y_p(^4\mathrm{\hat He}) \approx 
75\%$, much higher than in the SM. As we will show below, this has 
important consequences for LSS formation.

%%%%%%%%%%%%%%%%%%%%%%%%%%%%%%%%%%%%%%%
\subsection{Recombination}\label{s.rec}

When the temperature in the twin sector becomes much lower than the 
binding energy of mirror atoms, twin electrons $\hat{e}^-$ start to combine 
with twin hydrogen and helium nuclei into neutral bound states. This 
recombination process terminates the acoustic oscillations in the twin 
sector, and plays an important role in structure formation.

\subsubsection{\it Recombination in the SM}

Before considering recombination in the twin sector, it is helpful to 
first recall how this process occurs in the SM. Hydrogen provides the 
dominant contribution to the matter density in the SM. Therefore, in our 
analysis, we neglect the effects of helium, which are subdominant. The 
primary contribution to the recombination of hydrogen arises from the 
reaction $e^-+p\to \text{\text{H}}(n\geq 2)+\gamma$, followed by the 
decay from the excited state down to the H$(1s)$ state, rather than from 
direct capture to the ground state~\cite{Gorbunov:2011zz, 
Dodelson:2003ft,recomb}. This is because the direct capture of an 
electron into the H$(1s)$ state results in the emission of a hard photon 
that quickly ionizes a neighboring atom in the ground state, and 
therefore gives no net contribution to the recombination 
process.\footnote{The ionization cross section near threshold to the 
$1s$ atom is $\sim 10^7$ times larger than the Thomson cross section, 
which allows no opportunity to lose the photon energy before ionizing 
another $1s$ atom. Hubble expansion is also insufficient to redshift the 
photons arising from this transition to a low enough energy 
\cite{Gorbunov:2011zz}. Since both the ionization cross section and 
Thomson cross section are proportional to $m_e^{-2}$, the relative sizes 
of the two scattering processes remain the same in the MTH case.} We can 
simplify the process by considering just three electron states: ionized 
electrons, electrons in the $n=2$ state, and electrons in the ground 
state. Recombination then arises from the capture of an ionized electron 
into the $n=2$ state, followed by the de-excitation of the $n=2$ 
electron down to to $1s$, either through two photon emission, $2s\to 
1s+2\gamma$, or through Lyman-$\alpha$ decay, $2p\to 1s+\gamma$. In the 
case of de-excitation through Lyman-$\alpha$ decay, a net contribution 
to recombination only arises if the photon loses energy to redshift 
before colliding with another $1s$ electron.

We denote the ionized fraction of $e$ as $\chi_e \equiv 
n_e/n_{\text{H,tot}} = n_p/n_{\text{H,tot}}$, where $n_{\text{H,tot}} = 
n_p+n_{\text{\text{H}}(1s)}$ is the sum of both neutral and ionized 
hydrogen, $n_{\text{H,tot}}=8.6\times 10^{-6}\Omega_bh^2a^{-3}$cm$^{-3}$, and we have made use of the fact that helium has already recombined.
The Boltzmann equation for $\chi_e$ takes the form~\cite{Dodelson:2003ft,recomb,Kaplan:2009de},
 \begin{eqnarray}
 \label{eq:ionization}
\frac{d\chi_e}{dt}&=&-\alpha^{(2)}n_{\text{H,tot}}\,\chi_e^2+\beta\chi_2
\ ,
\\
\label{eq:ionizationdefn}
\chi_i&\equiv&\frac{n_{\text{\text{H}}(n=i)}}{n_{\text{H,tot}}},\quad\beta\equiv\frac{\alpha^{(2)}}{4}\left(\frac{m_eT}{2\pi}\right)^{3/2}e^{-\epsilon_0/4T} \ .
 \end{eqnarray}
 Here $\epsilon_0$ denotes the ground state energy of hydrogen, $13.6$ 
eV. In Eq.~(\ref{eq:ionization}), the first term on the right corresponds to 
the capture of ionized electrons into the $n=2$ state. Since the excited 
states of hydrogen are in thermal equilibrium, and the energy splitting 
between the $2s$ and $2p$ states $\sim\alpha_{em}^2\epsilon_0$ is much 
lower than the recombination temperature ($\sim \epsilon_0/10$), we do 
not distinguish between the two $n=2$ states. Therefore $\chi_2$ 
includes the contributions from both these states.

The second term on the right in Eq.~(\ref{eq:ionization}) corresponds to 
the ionization of the $n=2$ state, which releases electrons back into 
the ionized state. Both terms on the right in Eq.~(\ref{eq:ionization}) 
depend on the recombination cross section to the $n=2$ state, which can 
be approximated as \cite{1978ppim.book.....S,Ma:1995ey}
 \begin{equation}\label{eq:recoeff}
\alpha^{(2)}=0.448\frac{64\pi}{\sqrt{27\pi}}\frac{\alpha^2}{m_e^2}\left(\frac{\epsilon_0}{T}\right)^{1/2}\ln\left(\frac{\epsilon_0}{T}\right) \; ,
 \end{equation}
 for a general mass and coupling of a hydrogen-like atom.\footnote{As 
pointed out in Ref.~\cite{CyrRacine:2012fz}, Eq.~(\ref{eq:recoeff}) may 
not be a good approximation when the dark radiation temperature is 
higher or much lower than the binding energy, or when the dark proton is 
much colder than dark radiation. For the range that we are interested 
in, $3\leq (v_B/v_A)\leq5$, the relative sizes of the twin recombination 
temperature and twin particle masses is not very different from the SM, 
and so the twin electron remains in thermal equilibrium with twin 
photon. We therefore expect Eq.~(\ref{eq:recoeff}) to provide a good 
approximation in our case.} The net rate of production of $n=2$ 
hydrogen atoms is given by the equation
 \begin{equation}
\frac{d\chi_2}{dt}=\alpha^{(2)}n_{\text{H,tot}}\,\chi_e^2-\beta\chi_2-\left(\Lambda_{2\gamma}+\frac{H\omega^3_{\text{Ly}\alpha}}{\pi^2n_{\text{H,tot}}\,\chi_1}\right)\left(\frac{\chi_2}{4}-\chi_1e^{-\omega_{\text{Ly}\alpha}/T}\right).
 \end{equation}
 Here $\Lambda_{2\gamma}=8.227$ sec$^{-1}$ is the two photon decay rate, 
corresponding to the transition $2s\to 1s+2\gamma$, where neither photon 
has enough energy to excite a ground state hydrogen atom. The 
de-excitation can also come from the Lyman-$\alpha$ decay $2p\to 
1s+\gamma$, provided the redshift of the Lyman-$\alpha$ photon due to 
the expansion rate $H$ is faster than the re-absorption from the $n=1$ 
state determined by 
$(n_{\text{H,tot}}\,\chi_1\,\omega^{-3}_{\text{Ly}\alpha})$, where 
$\omega_{\text{Ly}\alpha}=3\epsilon_0/4$ is the energy of Lyman-$\alpha$ 
transition. For both these de-excitation processes, we have included the 
detailed balance correction corresponding to the reverse processes that 
arise from thermal excitation by background photons.

 When the production and destruction of the $n=2$ state is in 
equilibrium, $\frac{d\chi_2}{dt}=0$, we have
 \begin{equation}
\chi_2=4\frac{\alpha^{(2)}n_{\text{H,tot}}\,\chi_e^2+(\Lambda_{2\gamma}+\Lambda_{\alpha})\chi_1e^{-\omega_{\text{Ly}\alpha}/T}}{\Lambda_{2\gamma}+\Lambda_{\alpha}+4\beta},\quad\Lambda_{\alpha}=\frac{H(3\epsilon_0)^3}{(8\pi)^2n_{\text{H,tot}}\,\chi_1}.
 \end{equation}
 $\Lambda_{\alpha}\simeq 10$ s$^{-1}$ in the SM, which relates to the 
decay rate of $2p$ state by Lyman-$\alpha$ emission. The net rate of 
electron ionization in Eq.~(\ref{eq:ionization}) is
 \begin{equation}\label{eq:recomb}
\frac{d\chi_e}{dt}=-\frac{\Lambda_{\alpha}+\Lambda_{2\gamma}}{\Lambda_{\alpha}+\Lambda_{2\gamma}+4\beta}\,\alpha^{(2)}\left[n_{\text{H,tot}}\,\chi_e^2-(1-\chi_e)\left(\frac{m_eT}{2\pi}\right)^{3/2}e^{-\epsilon_0/T}\right].
 \end{equation}
 Here we have used the fact that $\chi_2\ll\chi_1$, which follows from 
detailed balance $\chi_2\simeq 
4\chi_1\exp(-\omega_{\text{Ly}\alpha}/T)$, to write $\chi_1\simeq 
1-\chi_e$. The ratio of the $\Lambda$ terms in front is of Eq. 
(\ref{eq:recomb}) is the Peebles correction, and the terms inside the 
square bracket correspond to those in the Saha equation that is derived from 
the thermal equilibrium of $p^++e^-\leftrightarrow \text{H}(1s)+$ 
photons
 \begin{equation}\label{eq:sahaH}
\frac{n_p\,n_e}{n_{\text{H}}(1s)}=\frac{n_{\text{H},\text{tot}}\,\chi_{e}^2}{1-\chi_{e}}= \left(\frac{m_{e}\,T}{2\pi}\right)^{3/2}e^{-\epsilon_0/T}.
 \end{equation}
 Although the Saha equation does not predict the correct relic abundance 
of electrons, it does approximate the starting point of recombination to 
a $15\%$ level precision.\footnote{For example, the Peebles equation 
predicts that half of the available hydrogen recombines by redshift 
$z\approx1200$, while the Saha equation predicts $z\approx 1400$.} We 
will use Eq.~(\ref{eq:recomb}) as the basis of our analysis for both SM 
and mirror hydrogen.

\subsubsection{\it Recombination in the Mirror Sector}

In order to generalize the analysis above to the case of twin hydrogen, 
we rescale the binding energy, mass, Lyman-$\alpha$ transition energy 
and temperature to their values in the mirror sector,
 \begin{equation}
\epsilon_0\to\hat{\epsilon}_0=\frac{v_B}{v_A}\epsilon_0,\quad m_e\to m_{\hat{e}}=\frac{v_B}{v_A}\,m_e,\quad\omega_{\text{Ly}\alpha}\to\hat{\omega}_{\text{Ly}\alpha}=\frac{v_B}{v_A}\,\omega_{\text{Ly}\alpha},\quad T\to \hat{T}=r_T\,T, 
 \end{equation}
 where $r_T$ is defined in Eq.~(\ref{eq:rT}). Since the Hubble expansion 
is mainly driven by the SM energy density, the expansion rate remains 
a function of SM temperature $H=H(T)$. Based on the fraction of twin 
hydrogen density $r_{\hat{\text{H}}}$ we use, the number density 
$n_{\text{H,tot}}$ is changed to the mirror density
 \begin{equation}
n_{\hat{\text{\text{H}}}\text{,tot}}=\frac{\Omega_{\text{DM}}\,r_{\hat{\text{H}}}\,m_p}{(1-Y_p(^4\mathrm{He}))\,\Omega_b\,m_{\hat{p}}}\,n_{\text{H,tot}}.
 \end{equation} 
 The two photon transition rate of a twin atom is given by 
\cite{1951ApJ...114..407S,CyrRacine:2012fz}
 \begin{equation}
\Lambda_{2\hat{\gamma}}=\left(\frac{\hat{\alpha}_{\text{em}}}{\alpha_{\text{em}}}\right)^{6}\left(\frac{\hat{\epsilon}_0}{\epsilon_0}\right)\Lambda_{2\gamma}=\left(\frac{v_B}{v_A}\right)\Lambda_{2\gamma}.
 \end{equation}

\begin{figure}
\begin{center}
\includegraphics[height=6cm]{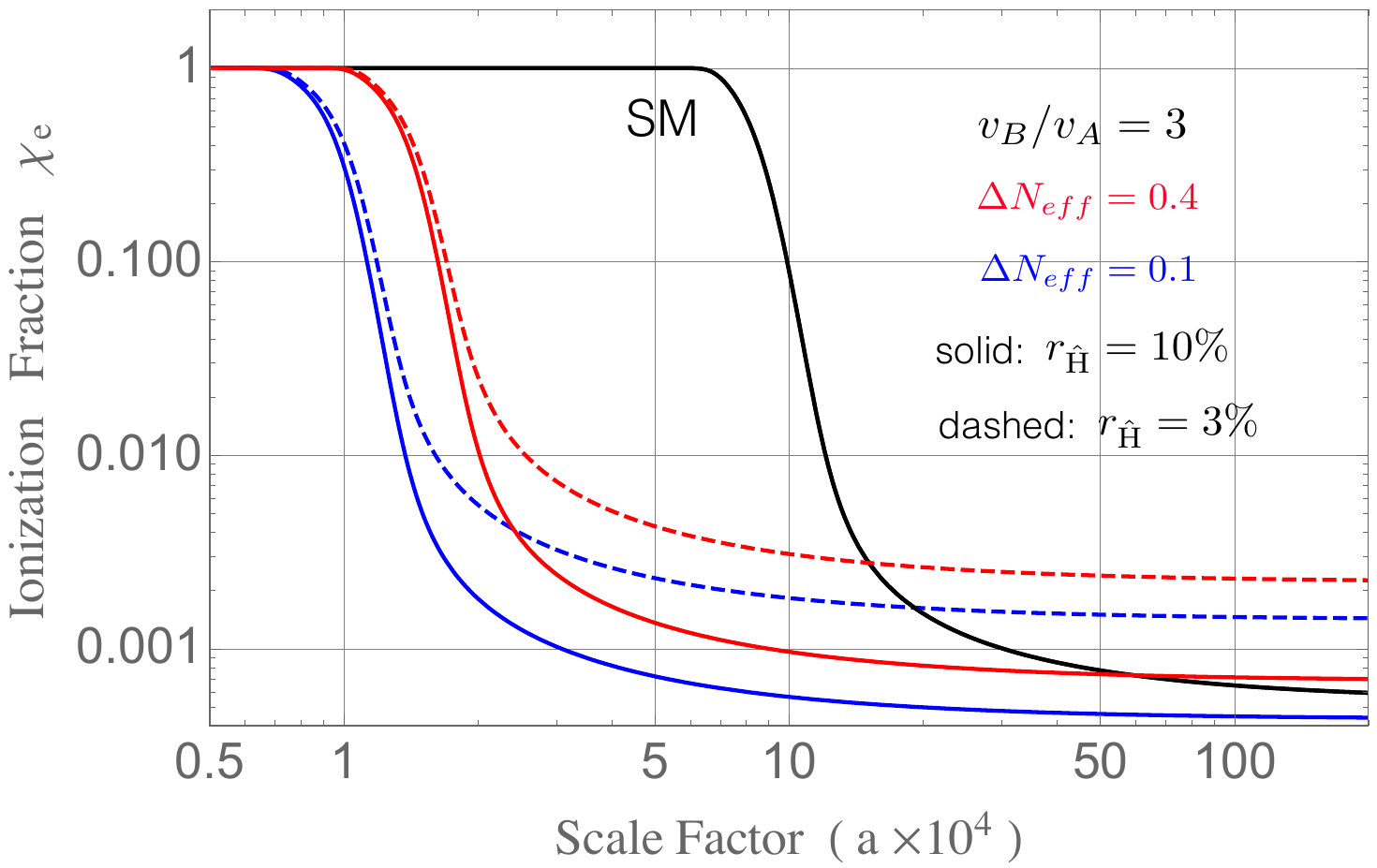}
\caption{The SM (blue) and Twin electron ionization fraction as a function of the scale factor. The MTH results are shown with different values of $r_{\hat{\text{H}}}=\Omega_{\hat{\text{H}}}/\Omega_{\text{DM}}$.}\label{fig:rec}
\end{center}
\end{figure}

We solve Eq.~(\ref{eq:recomb}) to obtain the ionization fraction of 
mirror hydrogen as a function of redshift for different values of 
$r_{\hat{\text{H}}}$ and $\Delta N_{eff}$. The results are plotted in 
Fig.~\ref{fig:rec}, with the SM ionization fraction provided for 
comparison. We see from the plot that larger $\Delta N_{eff}$, which 
corresponds to a higher twin sector temperature, is associated with 
later recombination.  A larger mirror hydrogen density and a lower twin 
temperature result in a smaller value of $\chi_e$ at freeze out.

The ionized states of mirror helium, $\hat{\text{H}}\text{e}^{2+}$ and 
$\hat{\text{H}}\text{e}^+$, have ionization energies $54.4(v_B/v_A)$ and 
$24.6(v_B/v_A)$ eV respectively. For $v_B/v_A \geq 3$, the universe is 
deep in the radiation dominated era when the temperature in the twin 
sector becomes comparable to the binding energy of mirror helium. 
However, due to its relatively large mass fraction shown in 
Fig.~\ref{f.mirrornoverp} (right), mirror helium still plays an 
important role in the formation of LSS, and cannot be neglected in the 
TBAO study. Mirror helium stops oscillating with the twin plasma after 
$\hat{\text{H}}\text{e}^+$ recombination. As with helium in the SM, the 
recombination of $\hat{\text{H}}\text{e}^+$ proceeds through complicated 
transitions between a network of excited levels. Instead of studying 
this process in detail, we estimate the time scale of recombination 
using the Saha equation. This approximation is justified because, in the 
SM, the Saha equation is known to reproduce the timescale of 
$\text{He}^+$ recombination to a precision of order $25\%$. From the 
Saha equation we obtain the $\hat{\text{H}}\text{e}^+$ ionization 
fraction as
 \begin{equation}\label{eq:saha}
\frac{n_{\hat{\text{H}},\text{tot}}\,\chi_{\hat{e}}\,\chi_{\hat{\text{H}}\text{e}^+}}{1-\chi_{\hat{\text{H}}\text{e}^+}}= 4\left(\frac{m_{\hat{e}}\,r_TT}{2\pi}\right)^{3/2}e^{-24.6\,\text{eV}\left(\frac{v_B}{v_A}\right)/r_TT}.
 \end{equation}
 For $v_B/v_A=3$ and $\Delta N_{eff}$ ranging from $0.1$ to $0.4$, the 
ionization fraction $\chi_{\hat{\text{H}}\text{e}^+}$ drops to $1\%$ at 
scale factors ranging from $a=5\cdot 10^{-5}$ to $7\cdot 10^{-5}$, 
corresponding to conformal times from $15$ to $20\,h^{-1}$Mpc. This means 
the $\hat{\text{H}}\text{e}^+$ scattering in the mirror plasma is expected 
to modify matter density perturbations starting from wavenumbers $\gsim 
0.05\,h$Mpc$^{-1}$. We will include the effects of twin helium oscillations 
on the matter power spectrum in the following section.

%%%%%%%%%%%%%%%%%%%%%%%%%%%%%%%%%%%%
\section{LSS Signals}\label{s.cmblss}

Prior to twin recombination, oscillations in the mirror baryon-photon 
fluid suppress the growth of structure in the twin sector. In contrast 
to PAcDM \cite{Chacko:2016kgg,Buen-Abad:2017gxg,Raveri:2017jto} and 
non-Abelian dark matter \cite{Buen-Abad:2015ova}, which also exhibit 
dark matter-dark radiation scattering, the oscillations in the MTH stop 
at a much earlier time because of twin recombination. Consequently, 
neutral twin atoms still give a sizable contribution to the matter 
density perturbations, and TBAO leaves an interesting residual 
oscillation pattern in the matter power spectrum. The overall 
suppression of structure on scales that enter prior to recombination, 
and the oscillatory pattern in the matter power spectrum, are 
characteristic features, not just of the MTH framework, but of the 
larger class of mirror models. For earlier work on LSS in the context of 
mirror models, see 
\cite{Ignatiev:2003js,Berezhiani:2003wj,Ciarcelluti:2004ik,Ciarcelluti:2004ip,Foot:2012ai}. 
Detailed studies of LSS for the general case of atomic dark matter may 
be found in \cite{CyrRacine:2012fz,Cyr-Racine:2013fsa}.

To determine the size of the corrections to LSS, we solve a set of 
linearized Boltzmann equations for density perturbations and calculate 
the ratio of the matter power spectrum in the MTH \emph{relative to} 
$\Lambda$CDM+DR (which is just $\Lambda$CDM with some extra dark 
radiation included to adjust the value of $\Delta N_{eff}$). In the MTH 
model, we consider a simplified scenario that contains only CDM $\chi$, 
ionized twin baryons 
$\hat{b}=\{\hat{\text{H}},\,\hat{\text{H}}\text{e}\}$, massless twin 
photons $\hat{\gamma}$, and the SM photons $\gamma$ and protons $p$. Due 
to their small energy density, (twin) electrons are neglected, but their 
effects are implicitly included since they mediate the interactions 
between (twin) protons and (twin) photons. We work in the conformal 
Newtonian gauge
 \begin{equation}
d s^2 
= a^2(\tau) \, 
  \bigl[ -(1 + 2 \psi) d\tau^2 + (1 - 2\phi) \delta_{ij} d x^i d x^j \bigr] \,,
 \end{equation}
 where the fields $\psi$ and $\phi$ describe scalar perturbations on the 
background metric. They are determined by four scalar quantities 
associated with the perturbed energy-momentum tensor $\delta T_\mu^\nu$, 
namely, $\delta \equiv \delta \rho / \bar{\rho} = -\delta T_0^0 / 
\bar{\rho}$, $\delta P = \delta T_i^i / 3$, $\theta \equiv -\partial_i 
\delta T_0^i / (\bar{\rho} + \bar{P})$, and $\sigma \equiv 
-\hat{\partial}_i \hat{\partial}^j (\delta T^i_j - \delta P \, 
\delta^i_j) / (\bar{\rho} + \bar{P})$, where $\bar{\rho}$ and $\bar{P}$ 
are the unperturbed total energy density and pressure, and 
$\hat{\partial}^i \equiv \hat{\partial}_i \equiv \partial_i / 
\sqrt{\partial_j \partial_j}$. For each particle species $s$, we define 
$\delta_s \equiv \delta\rho_s / \bar{\rho}_s$, $\theta_s \equiv 
-\partial_i \delta T_{s0}^i / (\bar{\rho}_s + \bar{P}_s)$, etc. We can 
also re-express $\theta_s$ as the divergence of comoving 3-velocity, 
$\theta_s = \partial_i v^i_s$, where $v^i \equiv d x^i / d \tau$. For 
each $s$ we assume an equation of state of the form $P_s = w_s \rho_s$ 
with constant $w_s$, so the pressures and energy densities are not 
independent quantities. The total $\delta$ and $\theta$ are given in 
terms of the individual $\delta_s$ and $\theta_s$ as $\delta = \sum_s 
\bar{\rho}_s \delta_s / \bar{\rho}$ and $\theta = \sum_s (\bar{\rho}_s + 
\bar{P}_s) \theta_s / (\bar{\rho} + \bar{P})$.

To linear order in the perturbations, the evolution of the dominant, 
collisionless component of dark matter, $\chi$, is described by \cite{Ma:1995ey}
 \begin{eqnarray} \label{eq:evolution1}
\dot{\delta}_{\chi}=-\theta_{\chi}+3\dot{\phi},\qquad
\dot{\theta}_{\chi}=-\frac{\dot{a}}{a}\theta_{\chi}+k^2\psi.
 \end{eqnarray}
 Here the derivatives are with respect to conformal time 
$\frac{d}{d\tau}$. For the oscillating component, the equations for the twin baryons
 are  given by 
 \begin{eqnarray}
\dot{\delta}_{\hat{b}}&=&-\theta_{\hat{b}}+3\dot{\phi} \ ,\label{eq:evolution3}
\\
\dot{\theta}_{\hat{b}}&=&-\frac{\dot{a}}{a}\theta_{\hat{b}}%+c_s^2k^2\delta_{\hat{p}}
+\frac{4\rho_{\hat{\gamma}}}{3\rho_{\hat{b}}}a\,n_{\hat{e}^{\pm}}(a)\hat{\sigma}_{T}(\theta_{\hat{\gamma}}-\theta_{\hat{b}})+k^2\psi \ .
\label{eq:evolution4}
 \end{eqnarray}
The corresponding equations for the SM baryons take a similar form. We 
solve for the $\hat{\text{H}}\text{e}^+$ recombination time using 
Eq.~(\ref{eq:saha}) and approximate the process as a step function in 
scale factor. As in the SM, neutral twin helium remains tightly 
coupled to $\hat{\text{H}}^+$ even after $\hat{\text{H}}\text{e}$ 
recombination. Therefore we continue to have 
$\hat{b}=\hat{\text{H}}+\hat{\text{H}}\text{e}$ even after 
$\hat{\text{H}}\text{e}$ recombination. However, the number density of 
free mirror electrons is reduced after twin helium recombination. This 
affects the time at which the twin photons decouple from the twin 
baryons, resulting in a reduction in the suppression of the matter power 
spectrum as compared to the case when only hydrogen is present. 
Therefore the matter power spectrum is sensitive to the relative 
abundances of $\hat{\text{H}}$ and $\hat{\text{H}}\text{e}$. The term 
that contains the Thomson cross section $\hat{\sigma}_T=6.7\cdot 
10^{-25}(v_A/v_B)^2$ cm$^2$ captures the effect of 
$\hat{b}$-$\hat{\gamma}$ scattering in the twin sector. The number 
density of the ionized twin electrons can be approximated as 
$n_{\hat{e}^{\pm}}(a)=[\Omega_{\hat{\text{H}}}(a)+\frac{1}{4}\Omega_{\hat{\text{H}}\text{e}}(a)]\rho_c/m_{\hat{\text{H}}}$ 
before twin helium recombination and 
$n_{\hat{e}^{\pm}}(a)=\chi_{\hat{e}}(a)\,\Omega_{\hat{\text{H}}}(a)\rho_c/m_{\hat{\text{H}}}$ 
afterwards, where the ionized twin baryon densities are given by 
$\Omega_{\hat{b}}=r_{\text{all}}\Omega_{\text{DM}}$ and 
$\Omega_{\hat{\text{H}}}=r_{\hat{\text{H}}}\Omega_{\text{DM}}$, and the 
ionization function of the twin electron $\chi_{\hat{e}}(a)$ is 
calculated numerically using the procedure outlined in Sec.~\ref{s.rec}.

The twin photon perturbations, including higher modes in the Legendre 
polynomials, evolve as
 \begin{eqnarray}
\dot{\delta}_{\hat{\gamma}}&=&-\frac{4}{3}\theta_{\hat{\gamma}}+4\dot{\phi},\label{eq:evolution5}
\\
\dot{\theta}_{\hat{\gamma}}&=&k^2\left(\frac{1}{4}\delta_{\hat{\gamma}}-\frac{1}{2}F_{\hat{\gamma}2}\right)+an_{\hat{e}}\hat{\sigma}_{T}(\theta_{\hat{b}}-\theta_{\hat{\gamma}})+k^2\psi,\label{eq:evolution6}
\\
\dot{F}_{\hat{\gamma}2}&=&\frac{8}{15}\theta_{\hat{\gamma}}-\frac{3}{5}k F_{\hat{\gamma}3}-\frac{9}{10}an_{\hat{e}}\hat{\sigma}_TF_{\hat{\gamma}2},\label{eq:evolution7}
\\
\dot{F}_{\hat{\gamma}l}&=&\frac{k}{2l+1}\left[lF_{\hat{\gamma}(l-1)}-(l+1)F_{\hat{\gamma}(l+1)}\right]-an_{\hat{e}}\hat{\sigma}_TF_{\hat{\gamma}l},\quad l\geq3\label{eq:evolution8}
\\
\dot{F}_{\hat{\gamma}l_{\text{max}}}&=&kF_{\hat{\gamma}(l_{\text{max}}-1)}-\frac{l_{\text{max}}+1}{\tau}F_{\hat{\gamma}l_{\text{max}}}-an_{\hat{e}}\hat{\sigma}_TF_{\hat{\gamma}l_{\text{max}}}.\label{eq:evolution9}
\end{eqnarray}
 Here the $F_{\hat{\gamma}l}$ are related to the spatial variations in 
the density fluctuations in the twin photons, $\delta_{\hat{\gamma}} 
\equiv F_{\hat{\gamma}0}$, $\theta_{\hat{\gamma}} \equiv 
\frac{3}{4}kF_{\hat{\gamma}1}/4$, and the shear stress $\sigma \equiv 
\frac{1}{2}F_{\hat{\gamma}2}$. We truncate the Boltzmann hierarchy at 
order $l_{\text{max}}=5$, making use of the approximation outlined in 
Ref.~\cite{Ma:1995ey}. As shown in the power spectrum ratio plot 
Fig.~\ref{fig:THpratio} (upper left), the result including only up to 
$l_{\text{max}}=4$ (dotted) exhibits only a small deviation from 
$l_{\text{max}}=5$ (solid). Since our focus is on describing the 
clustering of twin baryons in this framework, we do not solve for the 
twin photon polarization.

 We take $\psi=-\phi$ in our analysis, ignoring a small correction 
arising from the presence of free streaming radiation.\footnote{In the 
presence of free streaming radiation, the superhorizon gravity 
perturbation can be written as $ 
\psi=-\left(1+\frac{2}{5}R_{FS}\right)\phi$ and 
$R_{FS}=\left[1+\frac{8}{7(N_{\nu}+\Delta 
N_{eff,\hat{\nu}})(T_{\nu}/T_{\gamma})^4}\right]^{-1}$. However, once 
the mode enters the horizon, the anisotropic stress associated with the 
free streaming radiation quickly decreases and $\psi$ approaches 
$-\phi$.} Gravity perturbations are sourced by the density fluctuations 
as described by the Einstein equation,
 \begin{equation}
k^2\psi+3\frac{\dot{a}}{a}\left(\dot{\psi}+\frac{\dot{a}}{a}\psi\right)=-\frac{a^2}{2M_{pl}^2}\sum_{i=\chi,\hat{b},\hat{\gamma},p,\gamma}\rho_{i}\,\delta_{i}.\label{eq:evolution10}
 \end{equation}
 For the initial conditions, the modes that enter before 
matter-radiation equality satisfy
 \begin{equation}
\delta_{\gamma,\hat{\gamma}}=\frac{4}{3}\delta_{\chi,\hat{b},p}=-2\psi,\quad\theta_{\gamma,\hat{\gamma},\chi,\hat{b},p}=\frac{k^2\eta}{2}\psi,
 \end{equation}
 while for those come in during the era of matter domination,
 \begin{equation}
\frac{3}{4}\delta_{\gamma,\hat{\gamma}}=\delta_{\chi,\hat{b},p}=-2\psi,\quad\theta_{\gamma,\hat{\gamma},\chi,\hat{b},p}=\frac{k^2\eta}{3}\psi.
 \end{equation}
 We set the initial values of the higher modes 
$F_{\hat{\gamma}\ell\,\geq2}=0$, since these higher angular modes 
quickly damp away when the Thompson scattering is large. We neglect the 
tilt in the primordial spectrum ($n_s=1$) and take a $k$-independent 
value of $\psi= 10^{-4}$. The final results are independent of the 
precise value of $\psi$ since we are interested in the ratio of the 
matter power spectra with and without the twin oscillations. In the 
numerical study, we choose the values $h=0.68$, $\Omega_{\gamma} h^2 = 
2.47 \times 10^{-5}$, $\Omega_{\Lambda} h^2 = 0.69$, $\Omega_{b} h^2 = 
2.2\times 10^{-2}$ and $\Omega_{\nu}=0.69 \Omega_{\gamma}$ 
\cite{Ade:2015xua}.

We take as input the parameters of the MTH model 
$(r_{{\text{all}}},\,v_B/v_A $ and $\Delta N_{eff})$. Once these numbers 
are fixed, the mirror hydrogen and helium density fractions 
$r_{\hat{\text{H}}}$ and $r_{\hat{\text{H}}\text{e}}$ in 
Fig.~\ref{f.mirrornoverp}, the ionization function in 
Fig.~\ref{fig:rec}, and the rate of Thompson scattering in the twin 
sector are all determined. After solving for the density perturbations 
$\delta_{\chi,\hat{b},p}$, we calculate the total matter perturbation 
$\delta_{tot}$ and determine the relative suppression of the matter 
power spectrum with respect to $\Lambda$CDM+DR as,
 \begin{equation}\label{eq:psupress}
\delta_{tot}(k)=\sum_{i=\chi,\hat{b},p}(\Omega_i/\Omega_m)\,\delta_{i}(k),\qquad\text{P.S.}\,\,\text{Ratio}(k)\equiv\frac{\delta_{tot}^2(k)\Big|_{\Lambda\text{CDM}+\text{MTH}}}{\delta_{tot}^2(k)\Big|_{\Lambda\text{CDM}+\text{DR}}}.
 \end{equation}

\begin{figure}
\begin{center}
\includegraphics[height=5cm]{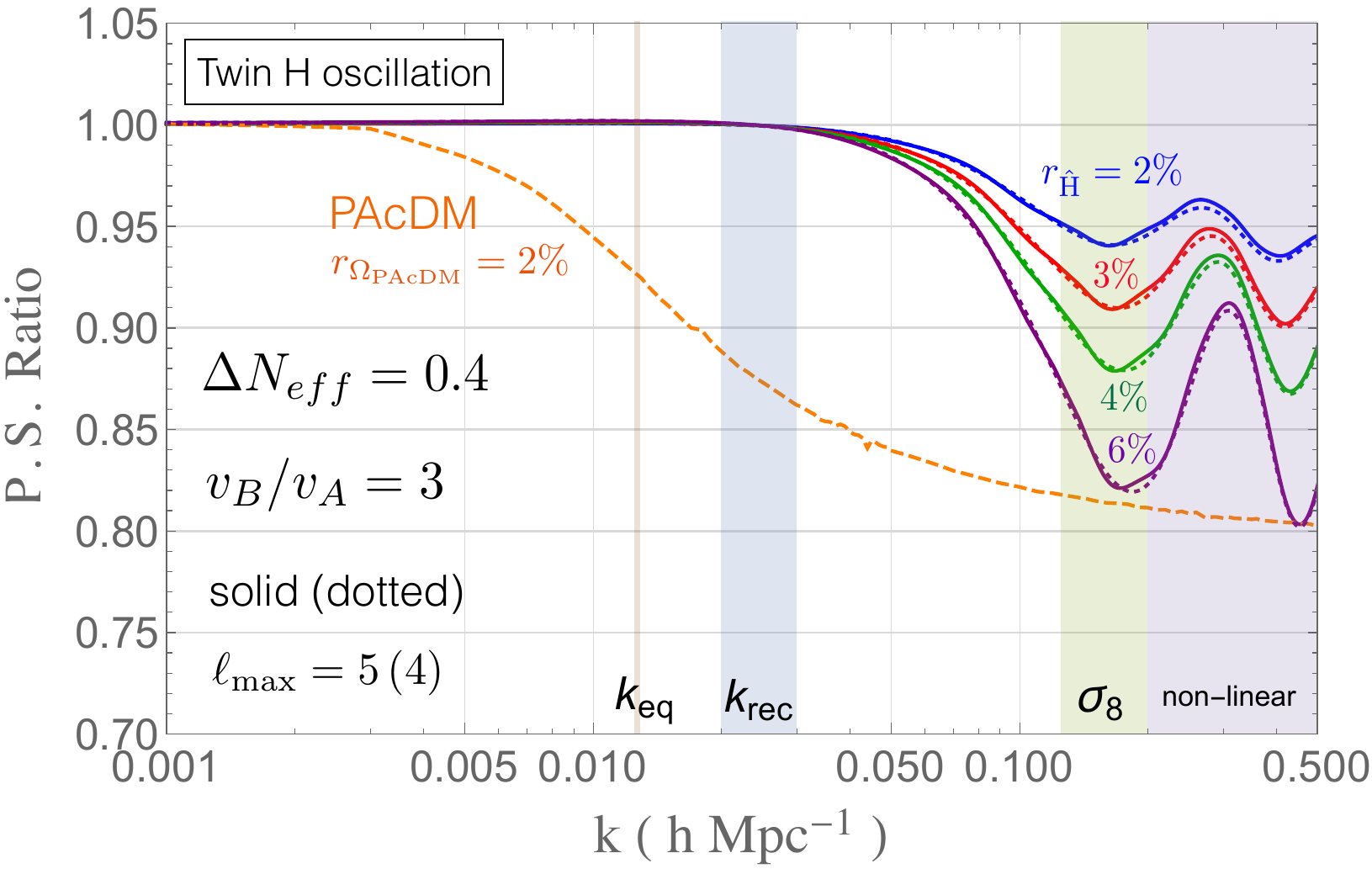}\,\,\includegraphics[height=5cm]{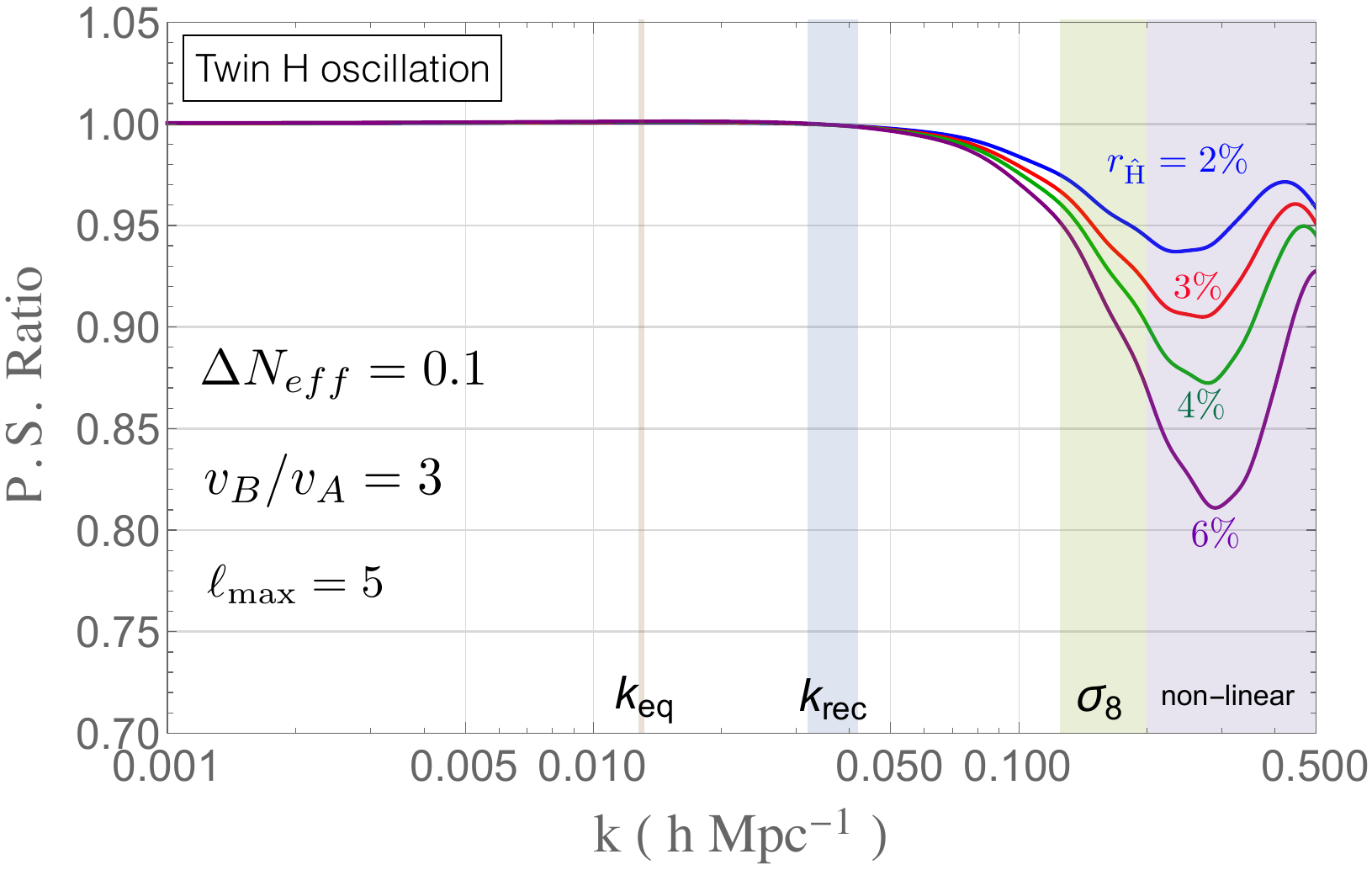}\\\includegraphics[height=5cm]{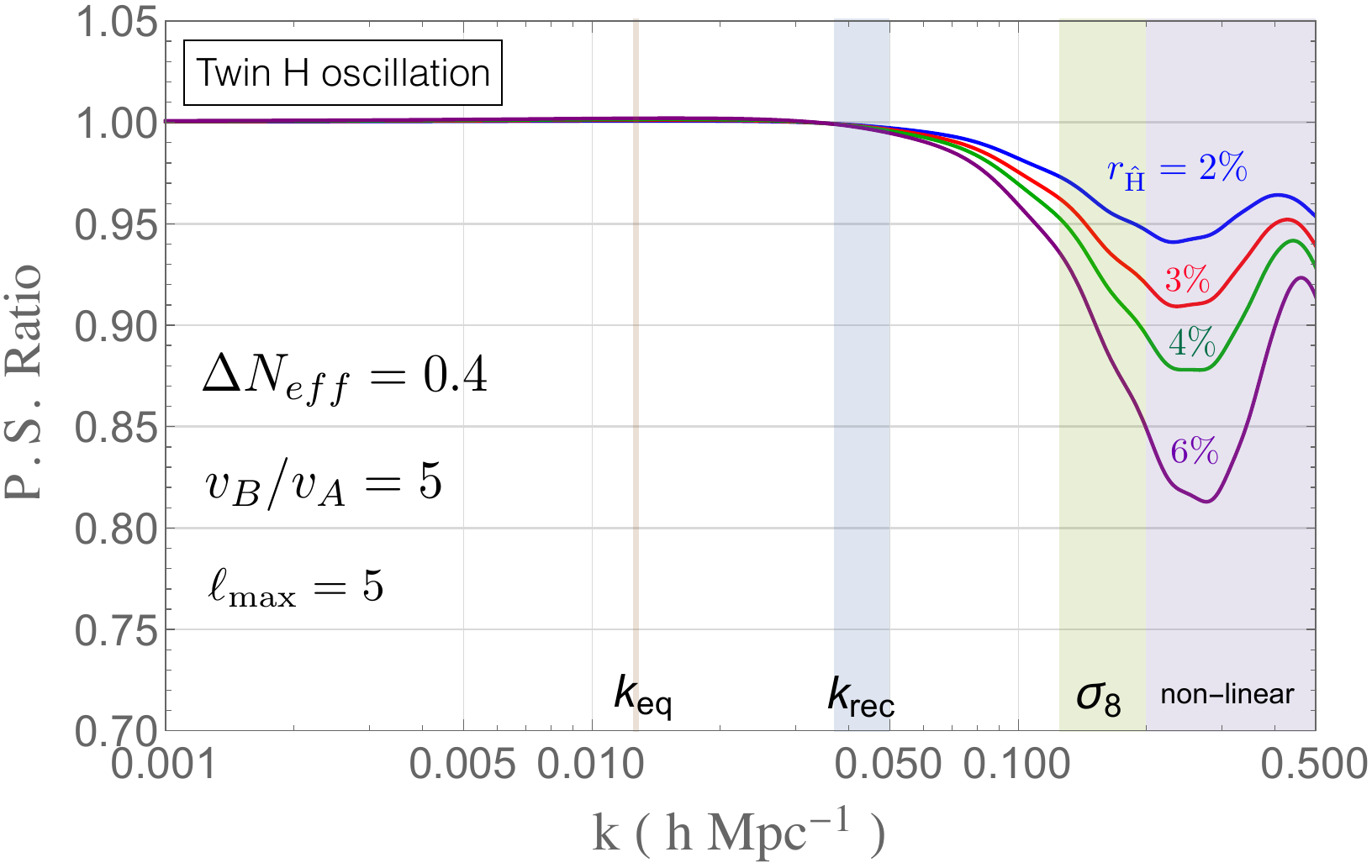}\,\,\includegraphics[height=5cm]{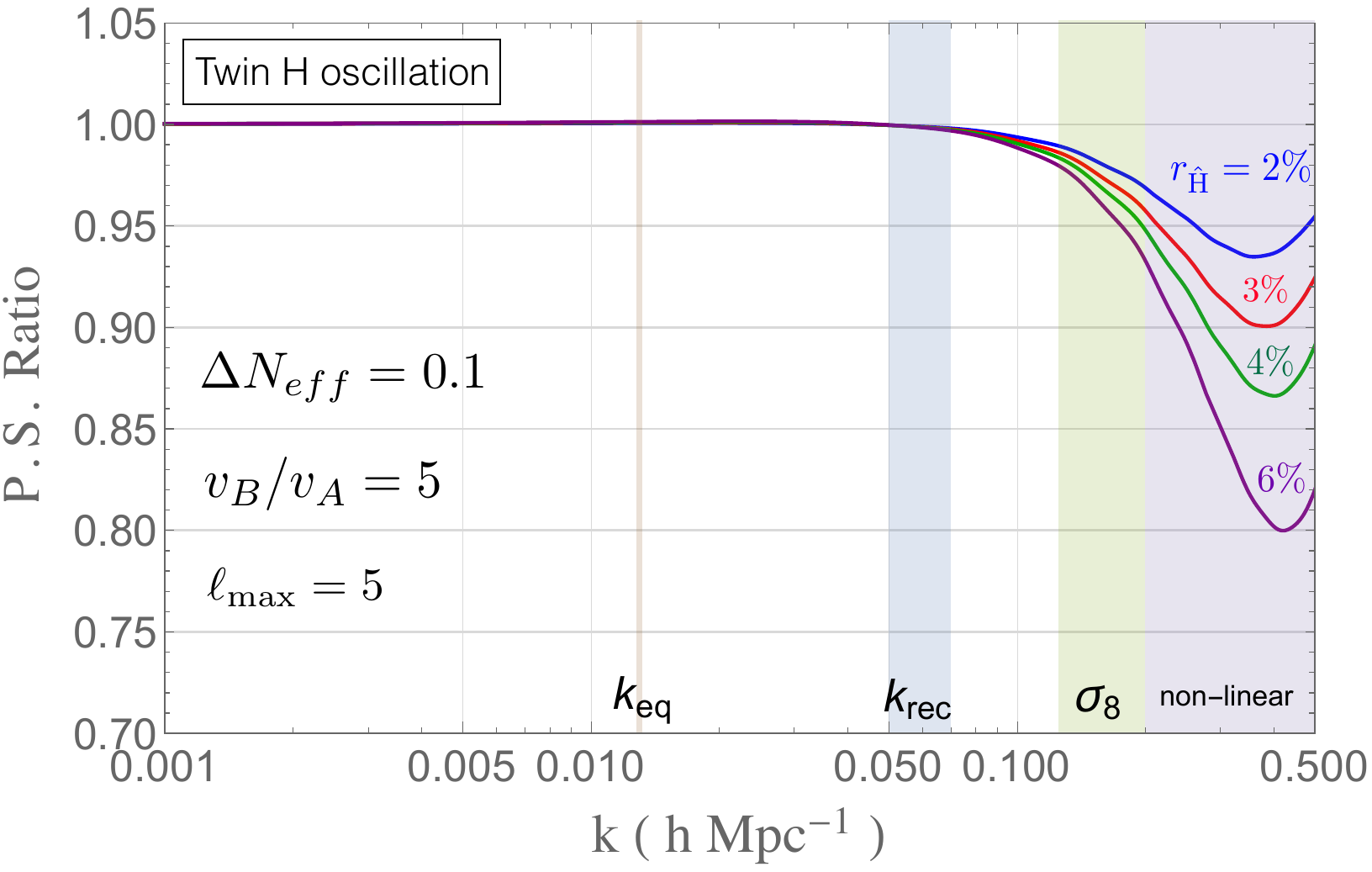}
\caption{Ratio of matter power spectrum between $\Lambda$CDM$+$MTH and 
$\Lambda$CDM$+$DR, both with $\Delta N_{eff}=0.4$ (left) and 0.1 (right) 
for $v_B/v_a$ = 3 (top) and 5 (bottom). Here we neglect the effect of 
twin Helium on the oscillation, assuming it is absent or has recombined 
much earlier. The orange dashed-dotted curve is from the PAcDM model 
\cite{Chacko:2016kgg}, in which a sub-component dark matter never 
decouples from the dark radiation scattering. For the MTH results, the 
solid (dotted) curves come from solutions of linearized Boltzmann 
equations discussed in Sec.~\ref{s.cmblss} with the inclusion of 
Boltzmann hierarchy modes up to $\ell_{\text{max}}=5$ (4). The vertical 
$k_{\text{rec}}$ band corresponds to the values of $k$ at the time of 
twin recombination over the range of $r_{\hat{\text{H}}}$ considered in the 
plot. The line $k_{eq}$ corresponds to the $k$ value of fluctuation mode that enters at matter-radiation equality.
} \label{fig:THpratio}
\end{center}
\end{figure}

Before studying the full problem that includes both twin helium and 
hydrogen, let us first consider the $\hat{H}$-only scenario to gain some physical 
intuition for the result.
In Fig.~\ref{fig:THpratio} we plot the 
results for four combinations of $(v_B/v_A$ and $\Delta N_{eff})$, 
assuming that $\hat{\text{H}}\text{e}$ is absent or has recombined much 
earlier so that $r_{\text{all}} = r_{\hat{\text{H}}}$. As compared 
to the PAcDM model, we see that the MTH exhibits a relatively sudden 
drop in the ratio of the matter power spectra. This happens at $k \approx 
0.04\,h$ Mpc$^{-1}$, which corresponds to the inverse of the conformal 
time $\tau_{rec}\approx 25\,h^{-1}$ Mpc at twin recombination. For larger 
$\Delta N_{eff}$, corresponding to a higher twin sector temperature, and 
smaller $v_B/v_A$, corresponding to a lower ionization energy, 
recombination happens later. The power spectrum oscillates around a 
constant suppression
 \begin{equation}
\text{P.S.}\,\,\text{Ratio}(k\gg\tau_{rec}^{-1})\simeq (1-r_{\hat{\text{H}}})^2.
 \end{equation}
 This scaling behavior is easy to understand. Twin recombination happens 
around the time of matter-radiation equality. Prior to this the density 
perturbations in cold dark matter grow logarithmically, 
$\delta_{\chi}(k)\simeq6\,\delta_{\chi,(i)}\ln\frac{k\tau}{\sqrt{3}}$ 
\cite{2011iteu.book.....G}. However, the twin protons and electrons 
undergo oscillations with the twin photons, leading to 
$\delta_{\hat{\text{H}}}\ll\delta_{\chi}$. It follows that the net matter 
power spectrum at the end of twin recombination is smaller than the 
$\Lambda$CDM result,
 \begin{equation}
\label{eq:rHapprx} 
\text{P.S.}\,\,\text{Ratio}(k\gg\tau_{rec}^{-1})\simeq\frac{\left[(1-r_{\hat{\text{H}}})\,6\,\delta_{\chi,(i)}\ln\left(\frac{k\tau_{rec}}{\sqrt{3}}\right)+r_{\hat{\text{H}}}\,\delta_{\hat{\text{H}}}\right]^2}{\left[6\,\delta_{\chi,(i)}\ln\left(\frac{k\tau_{rec}}{\sqrt{3}}\right)\right]^2}\approx 
(1-r_{\hat{\text{H}}})^2. 
 \end{equation} 
 After twin recombination, the dark matter density perturbations in both 
$\chi$ and $\hat{\text{H}}$ grow in the same way as for a single species of 
cold dark matter in $\Lambda$CDM, with the result that the ratio above 
is preserved.

 In addition to this overall suppression of the matter power spectrum, we 
see a residual oscillation pattern in the ratio of power spectra, with a 
period $\Delta k\approx 0.3\,h$Mpc$^{-1}$. There is also a subdominant 
oscillation with period $\Delta k\approx 0.06\,h$Mpc$^{-1}$ arising from 
interference with the SM BAO. In particular, the total density 
perturbation $\delta_{tot}$ in Eq.~(\ref{eq:psupress}) contains 
contributions from the density fluctuations in both the SM and twin 
proton components. Since the SM protons contribute more to the energy 
density than twin baryons, it is the SM BAO that generates the dominant 
oscillation pattern in $\delta^{2}_{tot}$. However, when considering the 
ratio of the two power spectra, this contribution cancels out so that 
the leading oscillatory effect arises from TBAO. As in SM BAO, the twin 
sector perturbations carry a $\cos(k\hat{r}_s)$ dependence in 
$\delta_{\hat{\text{H}}}$, with the sound horizon at recombination defined as
 \begin{equation}
\hat{r}_s\equiv\displaystyle{\int}_0^{\tau_{\text{twin\,rec}}}d\tau'\,c_s(\tau'),\qquad c_s(\tau)\equiv\left[3\,\left(1+\frac{3\,r_{\text{scatt}}\,\Omega_{\text{DM}}(\tau)}{4\,\Omega_{\hat{\gamma}}(\tau)}\right)\right]^{-\frac{1}{2}}.
 \end{equation}
 Here $r_{\text{scatt}}$ represents the dark matter mass fraction of the 
scattering twin ions. For $v_B/v_A=3$ and $\Delta N_{eff}=0.4$, the 
sound horizon at time of last scattering in the twin sector can be 
estimated as $\hat{r}_s\approx 20\,h^{-1}$Mpc. On the other hand, the 
oscillations in $\delta_p$ end at SM recombination, corresponding to a 
larger sound horizon $r_s\approx100\,h^{-1}$ Mpc. In 
Eq.~(\ref{eq:psupress}), the dominant contribution from TBAO shows up 
linearly in $\cos(k\hat{r}_s)$, corresponding to oscillations with 
period $\Delta k=2\pi/\hat{r}_s\approx 0.3h$ Mpc$^{-1}$. Interference 
between the SM and dark BAO also has an effect, but this is suppressed 
by an additional $r_{\hat{H}}\,\Omega_{b}/\Omega_{\text{DM}}$ in the 
ratio of power spectra as compared to TBAO, and has a period 
$\approx 1/5$ times shorter than twin oscillations.

\begin{figure}
\begin{center}
\includegraphics[height=4.9cm]{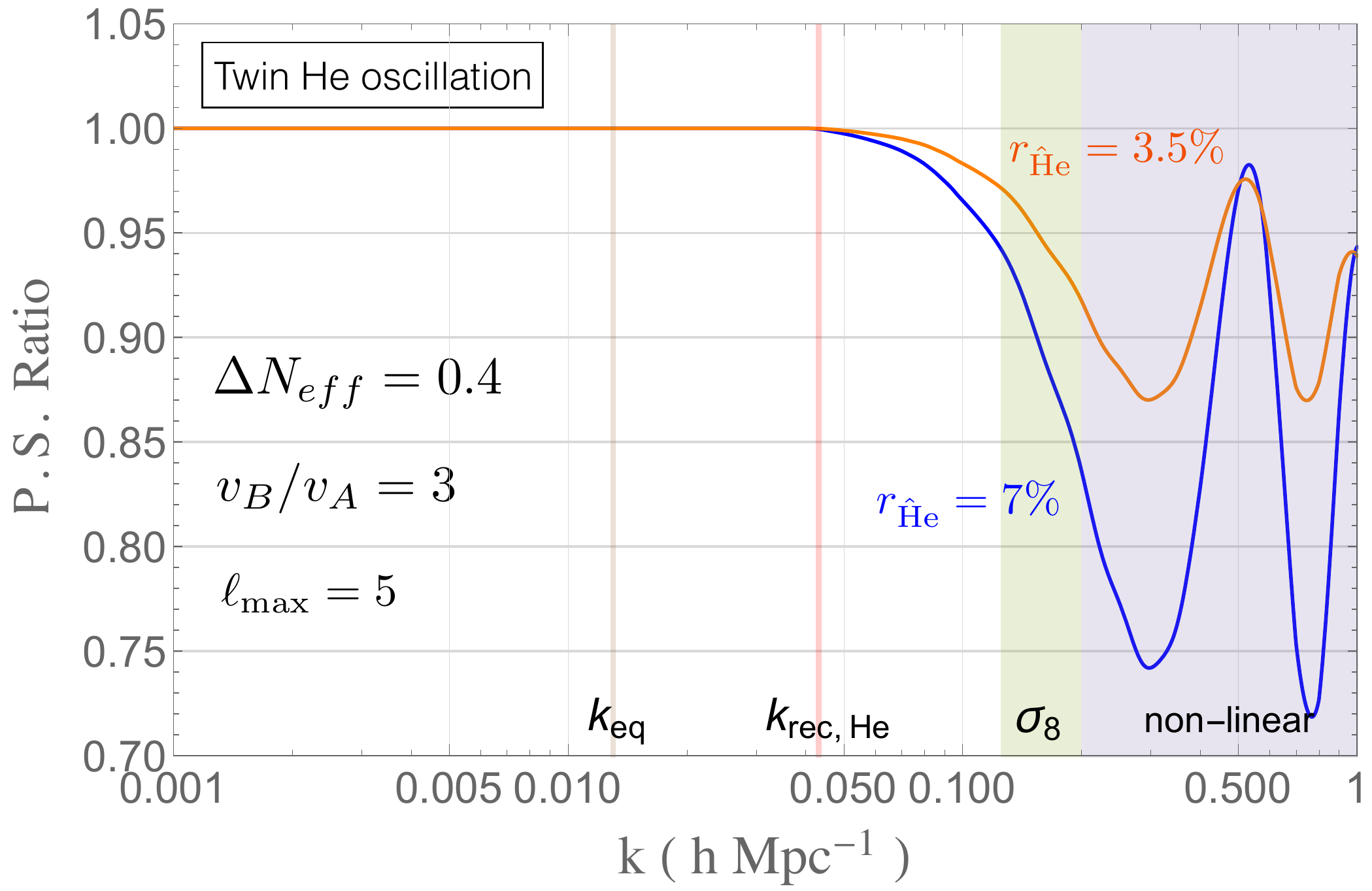}\,\,\,\,\,\includegraphics[height=4.9cm]{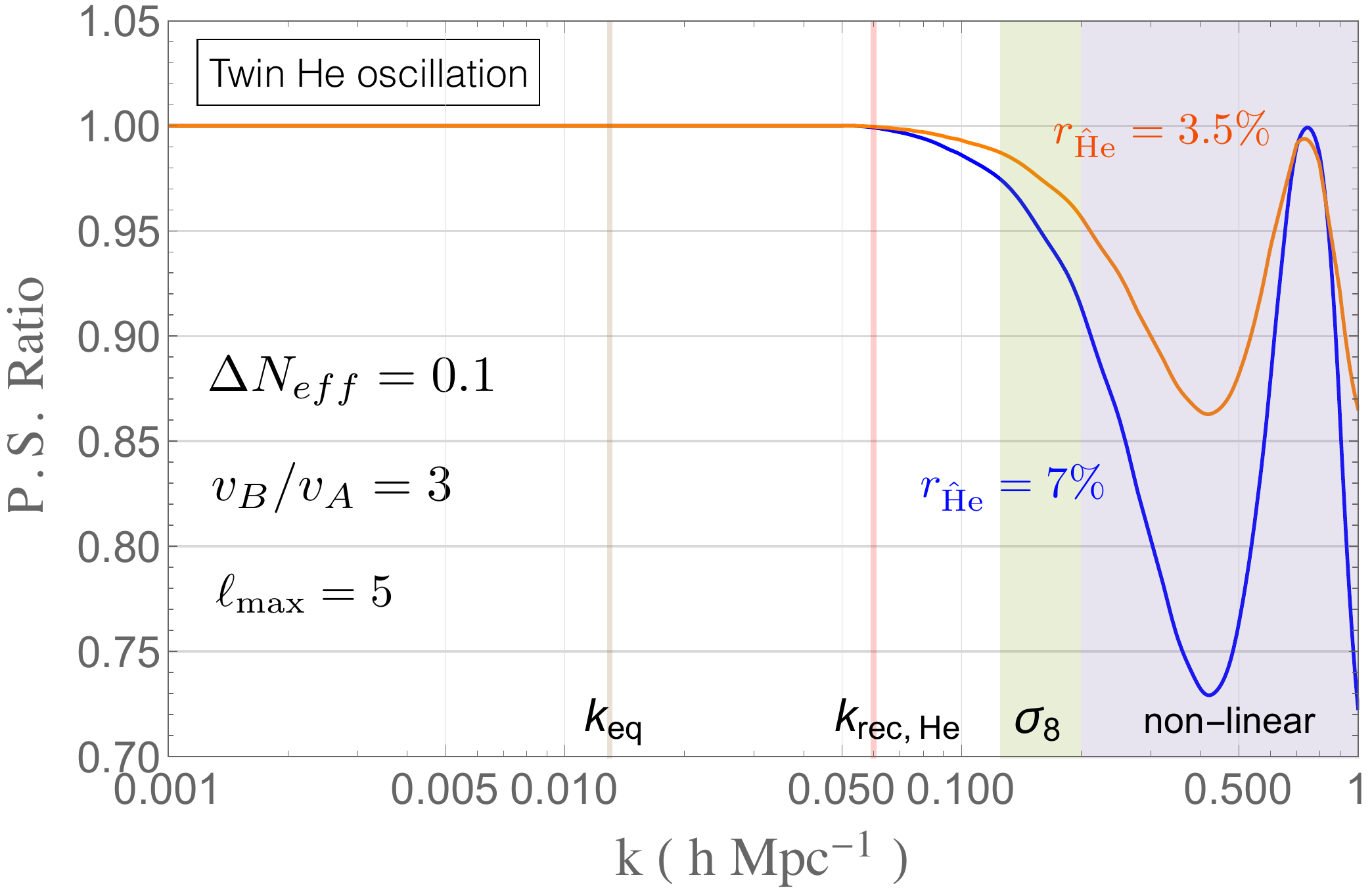}
 \caption{Ratio of matter power spectrum between $\Lambda$CDM$+$MTH and 
$\Lambda$CDM$+$DR, including only twin helium oscillations. 
$k_{\text{rec},\text{He}}$ corresponds to the time of twin helium 
recombination as obtained from the Saha equation Eq.~(\ref{eq:sahaH}).} 
\label{fig:THpratioHeonly}
 \end{center}
\end{figure}

We now consider the effects of mirror helium on LSS. In contrast to the 
SM, TBBN produces a much larger mass density of twin helium than twin 
hydrogen. Therefore twin helium plays an important role in the TBAO 
process, and cannot be neglected. From the Saha equation discussed in 
Sec.~\ref{s.rec}, $\hat{\text{H}}\text{e}^{+}$ recombines at conformal 
time $\approx 20\,h^{-1}$Mpc, and its scattering in the twin plasma 
leads to additional suppression in the power spectrum for $k\gsim 
0.05\,h$Mpc$^{-1}$. This means that $\hat{\text{H}}\text{e}^{+}$, while 
extremely important for the nonlinear regime with $k\gsim 
0.2\,h$Mpc$^{-1}$, also has a large impact on $\sigma_8$. To illustrate 
these effects, in 
Fig.~\ref{fig:THpratioHeonly} we show two representative matter
power spectra in which only the effects of mirror helium oscillations are 
included. The sound horizon at time of helium recombination has 
$\hat{r}_s\approx 17\,h^{-1}$Mpc, corresponding to oscillations with 
period $\Delta k \approx 0.6\,h$Mpc$^{-1}$, larger than in the case of 
hydrogen.

\begin{figure}
\begin{center}
\includegraphics[height=4.9cm]{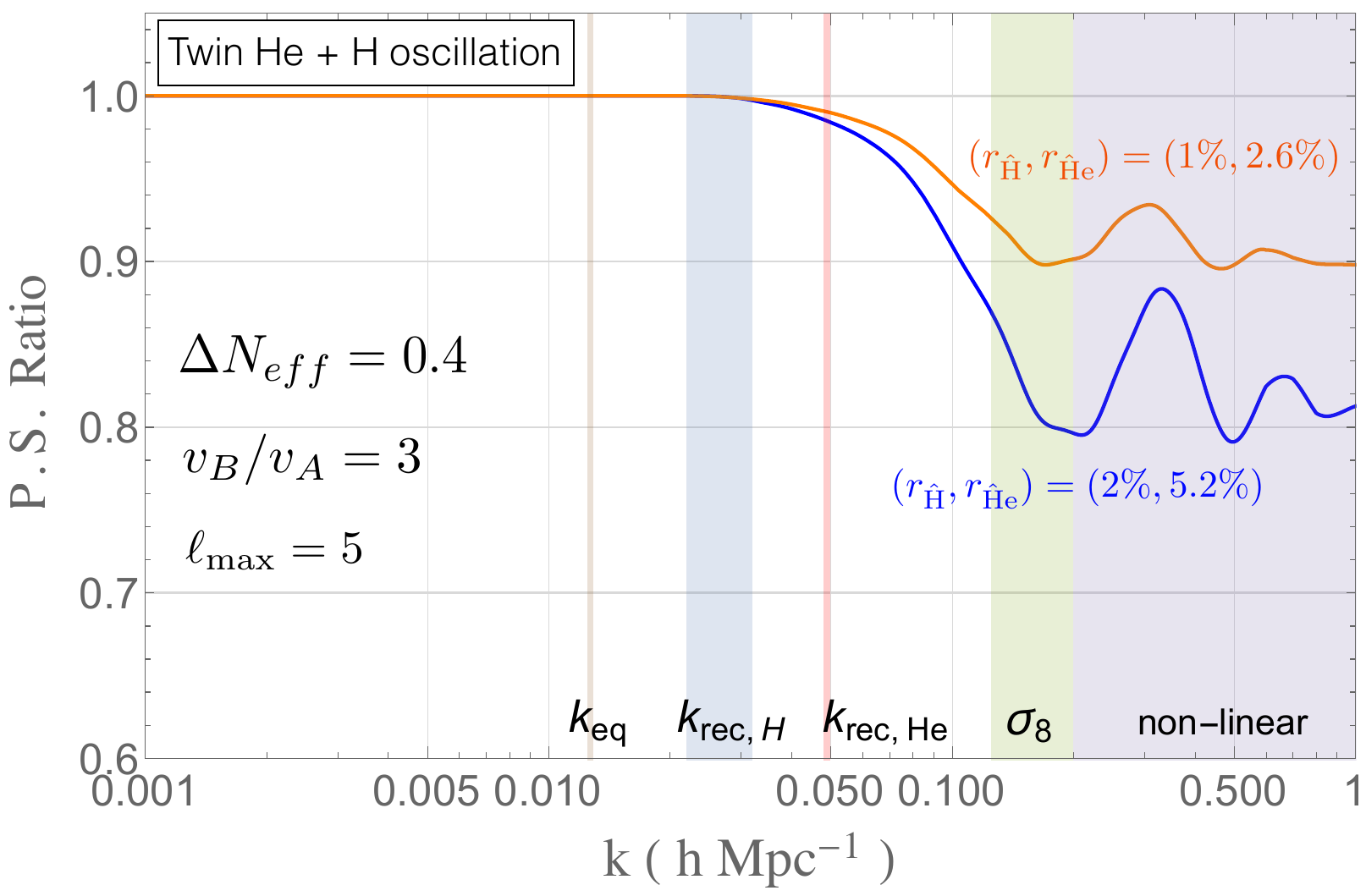}\,\,\,\,\,\includegraphics[height=4.9cm]{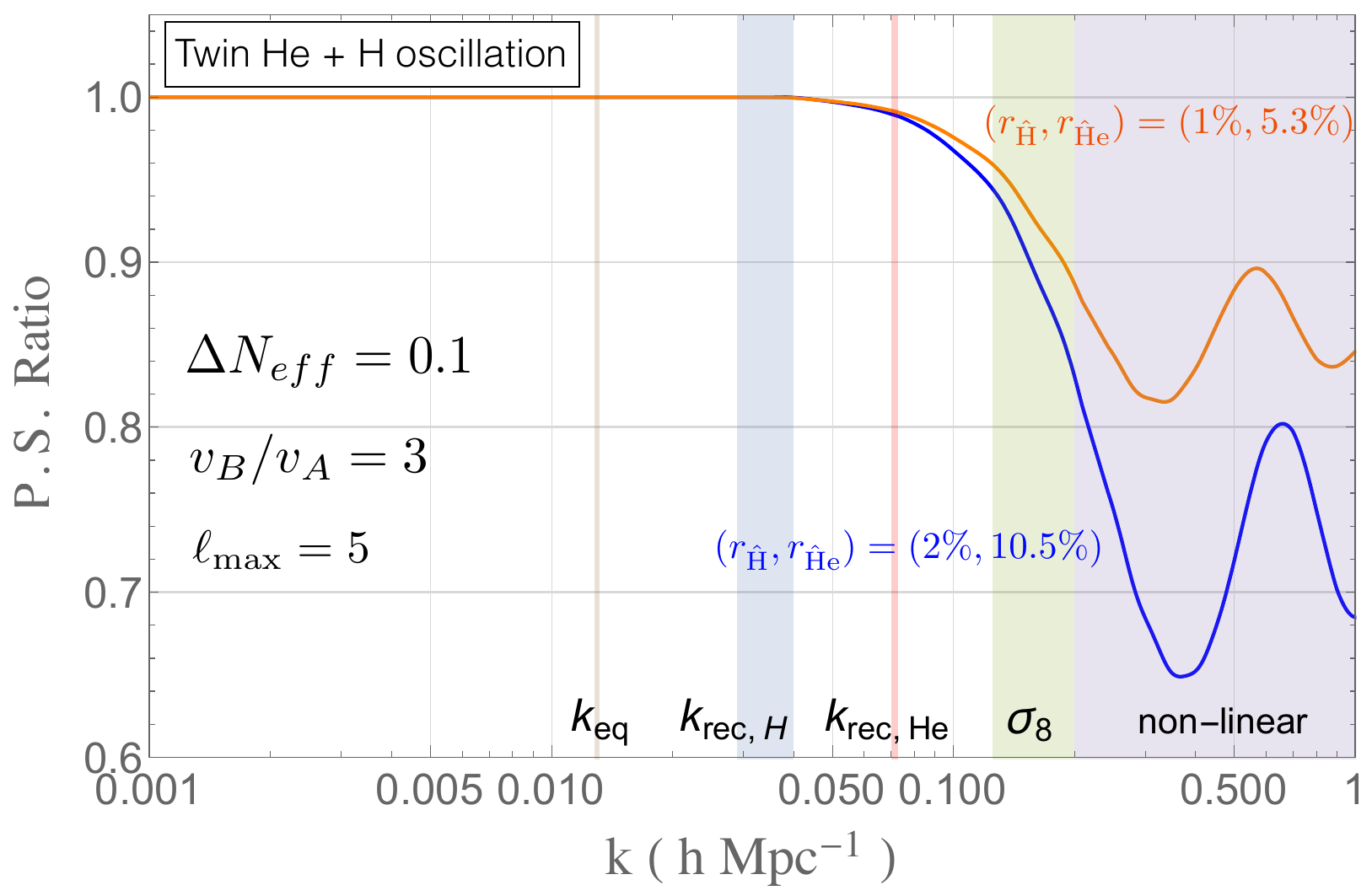}
\caption{Ratio of matter power spectrum between $\Lambda$CDM$+$MTH and 
$\Lambda$CDM$+$DR, including both the twin helium and hydrogen 
oscillations. The twin helium mass fractions are taken from 
Fig.~\ref{f.mirrornoverp}. $k_{\text{rec},\text{He}}$ corresponds to the 
time of twin helium recombination as obtained from the Saha equation 
Eq.~(\ref{eq:sahaH}).}
\label{fig:THpratioHe}
 \end{center}
\end{figure}

In Fig.~\ref{fig:THpratioHe} we present two examples of the power spectrum 
suppression, which take into account oscillations in both mirror 
hydrogen and helium. We take the twin helium density from 
Fig.~\ref{f.mirrornoverp} (right) for different values of 
$(\frac{v_B}{v_A}$ and $\Delta N_{eff})$ and terminate the helium 
oscillations at a recombination time when 
$\chi_{\hat{\text{H}}\text{e}^{+}}=1\%$ in Eq.~(\ref{eq:saha}).  
At later times
twin helium continues to behaves as an oscillating component of dark matter, 
but the number of free electrons
is reduced. As compared to the hydrogen-only 
scenario, the overall suppression in the power spectrum is now dominated 
by mirror helium. The magnitude of the suppression is still 
approximately given by Eq.~(\ref{eq:rHapprx}), but with the replacement 
$r_{\hat{\text{H}}}\to r_{\hat{\text{H}}}+r_{\hat{\text{H}}\text{e}}$. 
The sound horizon at the time of $\hat{\text{H}}^{+}$ recombination 
is given by $\hat{r}_s\approx 20\,h^{-1}$Mpc as shown in 
Fig.~\ref{fig:THpratioHe} (left), so the oscillation pattern exhibits a 
period which is similar to the hydrogen-only case. 
The small distortions in the curves arise from interference
between the mirror and SM baryon oscillations. For a 
smaller $\Delta N_{eff}$, corresponding to a lower temperature in the 
twin sector, the earlier freeze out of the TBBN processes results in 
more mirror helium. Furthermore, the lower temperature makes the mirror 
helium recombine earlier. For a given $r_{{\text{all}}}$ and $v_B/v_A $ 
this results in the same overall suppression of the power spectrum deep 
in the nonlinear regime, but a smaller reduction in $\sigma_8$.

 \begin{figure} 
\begin{center} 
\includegraphics[height=7cm]{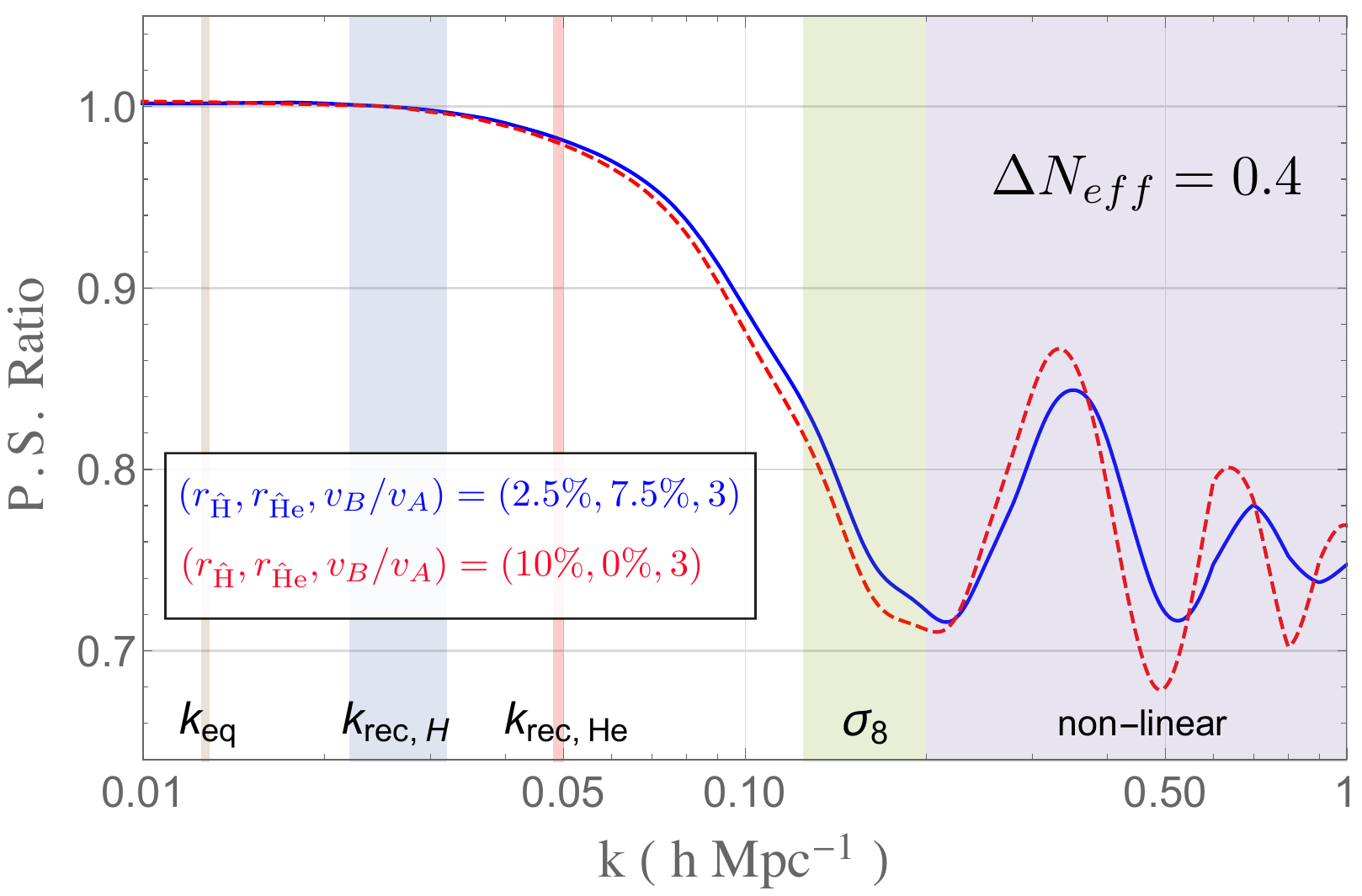}  
\caption{
Ratio of the matter power spectra between $\Lambda$CDM$+$MTH and 
$\Lambda$CDM$+$DR. Results are shown for two different abundances of 
mirror H and He, but for the same value of $v_B/v_A$.
The blue curve considers $r_{\rm all} = 0.1$ with 
$Y_{\hat{\rm H}{\rm e}}=0.75$, which is close to the value from the full 
MTH scenario studied in Sec.~\ref{s.BBN}. The red dashed curve 
corresponds to the hydrogen-only scenario, again with $r_{\rm all} = 
0.1$.  The two curves begin to exhibit percent level
differences for $k\gsim 0.1h$ Mpc$^{-1}$. However, since the blue and red 
curves correspond to the same $r_{\rm all}$ for the blue and red curves 
the average suppression for large $k$-modes is the same, as discussed in 
Eq.~(\ref{eq:rHapprx}).
 }
\label{fig:ADM2vs1samev} 
\end{center} 
 \end{figure}
 
In Fig.~\ref{fig:THpratioHe}, the power spectrum suppression for 
$k\lsim 0.1\,h$Mpc$^{-1}$ depends on the time scale of $\hat{\rm H}$ 
recombination. For a given $r_{\rm all}$, the relative densities of 
mirror hydrogen and helium determine the number of ionized electrons 
that survive after $\hat{\rm H}{\rm e}$ recombination. Consequently the 
time at which the acoustic oscillations in the mirror sector cease 
depends on the ratio $r_{\hat{\rm H}{\rm e}}/r_{\rm all}$.
Therefore the matter power spectrum is sensitive to the relative 
abundances of mirror hydrogen and helium, as shown in 
Fig.~\ref{fig:ADM2vs1samev}. This opens the door to the possibility 
of distinguishing the MTH universe from scenarios with a single species 
of dark atom. As shown in Fig.~\ref{fig:ADM2vs1}, although it is 
possible to match part of the $\hat{\text{H}}+\hat{\text{H}}\text{e}$ 
result (blue curve) using two different twin hydrogen-only (red dashed 
and green dotted) scenarios, the fit necessarily leaves residual 
differences with the MTH in either the linear regime, the nonlinear regime, 
or both.
If we hold $r_{\rm all}$ the same as in the MTH case but allow $v_B/v_A$ 
to float (the red curve), we can match the blue curve at low $k$, but 
find that percent level differences remain even in the linear regime 
near $k=0.2\,h$Mpc$^{-1}$. In the nonlinear regime, the differences 
between the red and blue curves are much larger, but the average 
suppression of the power spectrum in the two cases remains the same. If, 
however, we allow both $r_{\rm all}$ and $v_B/v_A$ to float (the green 
curve), it is possible to match the blue curve in the entire linear 
regime. However, there are still large differences between the shapes of 
the blue and green curves in the nonlinear regime, and even the average 
suppression of the power spectrum in the two cases is different. Future 
experiments from weak lensing data are expected to constrain the matter 
power spectrum for $k\lsim 0.5h$Mpc$^{-1}$ to percent level precision. 
Although the $k\gsim 0.5h$Mpc$^{-1}$ region is plagued by nonlinear 
effects, even here future measurements based on 21 cm tomography will 
probe the universe at much higher redshifts, $z\sim 100$, and therefore 
allow a detailed study of $\mathcal{O}(1)$ Mpc $k$-modes before they 
enter the nonlinear regime. Therefore, future observations of the matter 
power spectrum may be able to reveal the existence of more than one 
species of dark atom, which is a striking prediction of the MTH 
framework.

\begin{figure} 
\begin{center} 
\includegraphics[height=7cm]{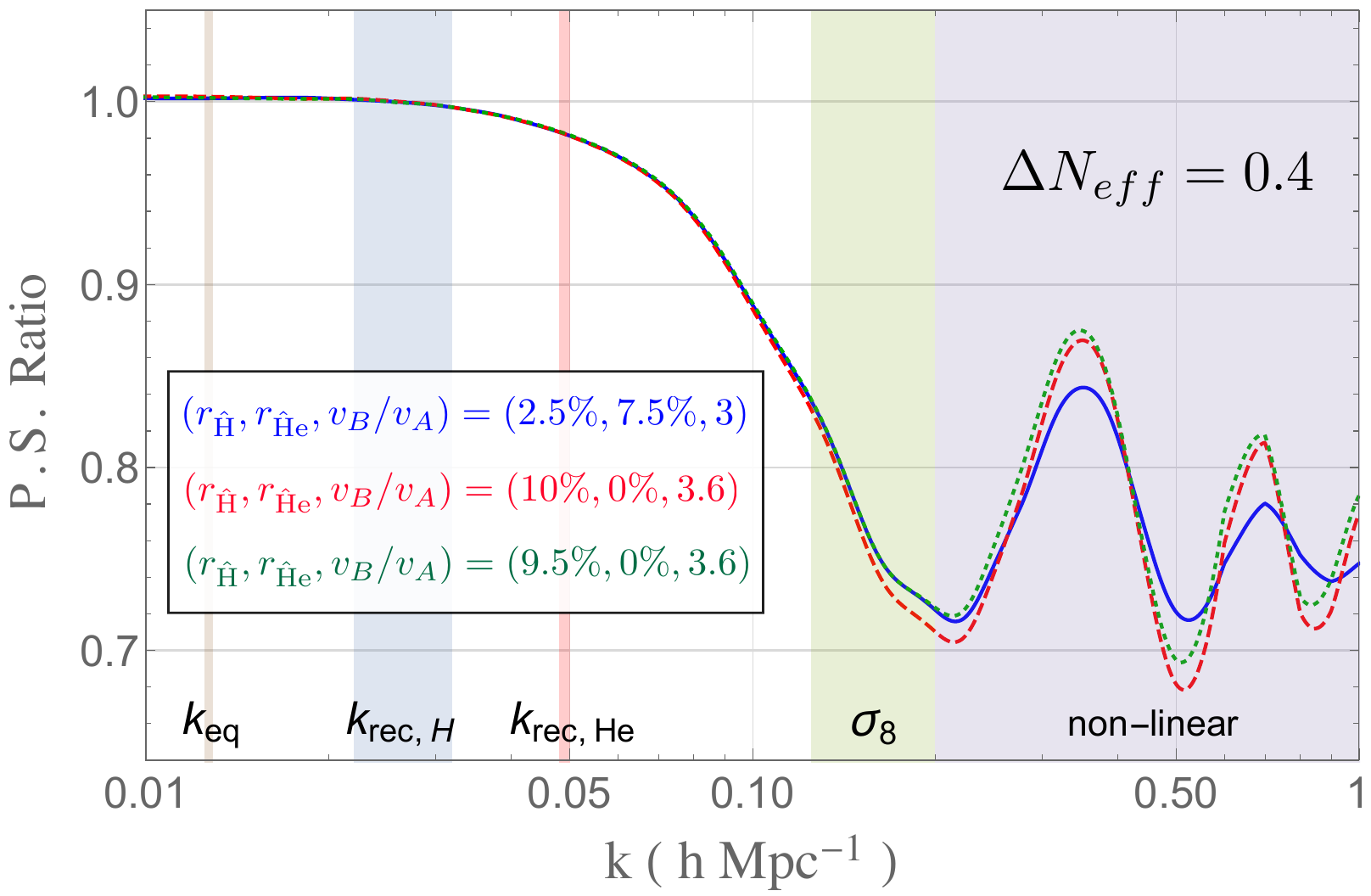}
\caption{Ratio of the matter power spectra between $\Lambda$CDM$+$MTH and 
$\Lambda$CDM$+$DR, for three different abundances of 
mirror H and He. The blue curve considers $r_{\rm all} = 0.1$ with 
$Y_{\hat{\rm H}{\rm e}}=0.75$, which is close to the value from the full 
MTH scenario studied in Sec.~\ref{s.BBN}. The red dashed curve 
corresponds to the hydrogen-only scenario, again with $r_{\rm all} = 
0.1$, but with a larger value of $v_b/v_A$ corresponding to a heavier 
mirror electron.  We see that this gives a good fit to the MTH result 
for $k\lsim 0.12\,h$Mpc$^{-1}$, but the two curves begin to exhibit 
differences in the $\sigma_8$ region. However, since the blue and red 
curves correspond to the same $r_{\rm all}$ for the blue and red curves 
the average suppression for large $k$-modes is the same, as discussed in 
Eq.~(\ref{eq:rHapprx}). The green dotted curve corresponds to values of 
$r_{\rm all}$ and $v_b/v_A$ that have been chosen to mimic as closely as 
possible the blue curve in the linear regime, $k\lsim 0.2\,h$Mpc$^{-1}$. 
However, since $r_{\rm all}$ differs from $0.1$, the average suppression 
for large $k$-modes deviates from the blue curve. The percent level 
differences between the blue and red curves at $k\approx 
0.2\,h$Mpc$^{-1}$ and between the blue and green curves at $k> 
0.2\,h$Mpc$^{-1}$ may allow future matter power spectrum measurements to 
distinguish the MTH from theories with a single species of dark atom.
 }
\label{fig:ADM2vs1} 
\end{center} 
 \end{figure}

If TBAO is to provide an explanation for the current $\approx 10\%$ 
discrepancy in $\sigma_8$ between the low redshift measurements and the 
Planck fit, the fractional contribution of twin hydrogen to the overall 
dark matter density is required to be $r_{\hat{\text{H}}}\approx 1\%$, 
corresponding to $r_\mathrm{all} \sim 3.6\%$, for $v_B/v_A= 3$ and $\Delta 
N_{eff}=0.4$. For these values of $r_{\mathrm{all}}$ and $v_B/v_A$, 
the superimposed oscillation of the modes with $k>0.2\,h$ Mpc$^{-1}$ 
leads to an extra $\approx 4\%$ variation of the power spectrum as a 
function of $k$. This is particularly significant keeping in mind the 
expected $\mathcal{O}(1)\%$ sensitivity of future LSS observations 
\cite{Archidiacono:2016lnv}, which will provide an observational channel to 
probe the MTH model. Nonlinear effects, however, become important for 
$k\gsim 0.2h$ Mpc$^{-1}$, and a detailed study in the higher $k$-mode 
region is required to determine the full oscillation pattern.
 \begin{figure} 
\begin{center} 
\includegraphics[height=8cm]{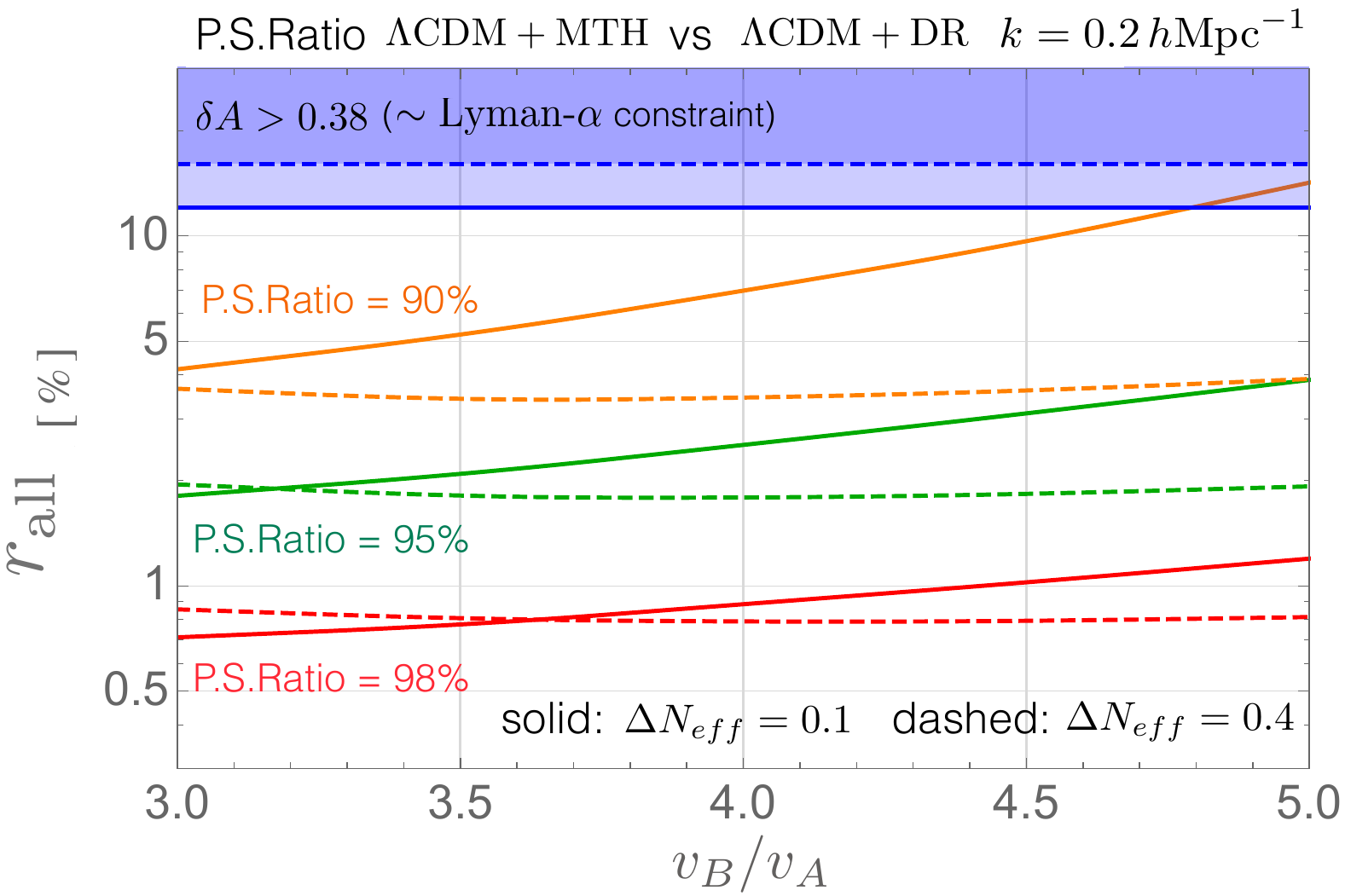} 
\caption{
 Estimation of projected constraints on the twin baryon mass fraction 
$r_{\mathrm{all}}$ from future LSS measurements, as a function of 
$v_B/v_A$. Next generation lensing measurements are expected to bound 
the power spectrum suppression at the few percent level 
\cite{Archidiacono:2016lnv}, so we show the constraints arising from the 
90, 95 and 98\% lower limits on the ratio of power spectra defined in 
Eq.~(\ref{eq:psupress}) as the orange, green and red curves. The ratio 
of the densities of twin hydrogen and helium are obtained from 
Fig.~\ref{f.mirrornoverp}. The region above the solid (dashed) blue line 
for $\Delta N_{eff}=0.1\,(0.4)$ represents an approximation to the 
Lyman-$\alpha$ constraint that excludes scenarios with $\delta A$, the 
deviation of the integrated power spectrum defined in 
Eq.~(\ref{eq:Aratio}), larger than $38\%$.
 }
\label{fig:LSSbound} 
\end{center} 
 \end{figure}

In the absence of a signal, future measurements of LSS will be able to 
place stringent limits on the energy density in twin baryons. In 
Fig.~\ref{fig:LSSbound} we show the upper bounds on the fractional 
contribution of twin baryons to the dark matter density, 
$r_{\text{all}}$, corresponding to different current and projected lower 
bounds on the matter power spectrum suppression, as defined in 
Eq.~(\ref{eq:psupress}). The effects of both twin helium and hydrogen 
are included. We present the results for $k=0.2\,h$Mpc$^{-1}$, which 
exhibits a significant suppression of the power spectrum, but for which 
nonlinear effects are still under theoretical control. Current 
measurements of the matter power spectrum allow a $\lsim 10\%$ deviation 
from the $\Lambda$CDM prediction at this scale, which constrains the 
fractional contribution of twin baryons to the dark matter density to 
lie in the range $r_{\text{all}}\lsim 4$-$10\%$, depending on the other 
parameters. Given the expected $\mathcal{O}(1)\%$ sensitivity of future 
LSS observations, if the observed result is fully consistent with the 
$\Lambda$CDM prediction, we expect to be able to constrain the density 
fraction to $r_{\text{all}}\lsim 1\%$. Since cooler dark radiation 
(corresponding to a smaller $\Delta N_{eff}$) and a heavier twin 
electron (larger $v_B/v_A$) lead to earlier recombination, the 
corrections to the power spectrum are smaller in this case. This 
explains the weaker bounds on $r_{\mathrm{all}}$ for the $\Delta 
N_{eff}=0.1$ case with large $v_B/v_A$.

The TBAO suppression persists out to $k>0.5\,h$ Mpc$^{-1}$. Then the 
Lyman-$\alpha$ observations, which probe the matter power spectrum on 
scales $0.5\,h^{-1}$ Mpc $<\lambda<100\,h^{-1}$ Mpc \cite{Viel:2013apy} 
at $z\approx 3$-$5$ can also be used to constrain $r_{\text{all}}$. A 
precise Lyman-$\alpha$ bound requires detailed N-body simulations of the 
MTH plasma, which are beyond the scope of this work. Here we only give a 
rough estimate of the current bound following the strategy adopted in 
Ref.~\cite{Murgia:2017lwo} for the warm dark matter (WDM) study, which 
constrains the integrated matter power spectrum over a range of wave 
numbers $0.5<k<20\,h$ Mpc$^{-1}$\footnote{The study is based on the 
assumption that the bias function $b(k)$ between the flux power spectrum 
$P(k)_F$ and the linear power spectrum $P(k)$, written as 
$P(k)_F=b^2(k)\,P(k)$, only differs a little between $\Lambda$CDM and 
the new model. The integration over $k$-modes is justified by the fact 
that velocities in the Intergalactic Medium tend to redistribute the 
power spectrum within a range of wave numbers in the probed region 
\cite{Gnedin:2001wg}.}. We define the ratio factor
 \begin{equation}\label{eq:Aratio}
\delta A\equiv\frac{A_{\Lambda\text{CDM+DR}}-A_{\Lambda\text{CDM+MTH}}}{A_{\Lambda\text{CDM+DR}}},\quad A_{\Lambda\text{CDM+MTH}}=\displaystyle{\int}_{k_{\text{min}}}^{k_{\text{max}}}dk\,\,\text{P.S.Ratio}(k)
 \end{equation}
 and determine the result for redshift $z=3$. By construction, 
$A_{\Lambda\text{CDM+DR}}=k_{\text{max}}-k_{\text{min}}$. We take 
$\delta A<0.38$ from Ref.~\cite{Murgia:2017lwo} as a 2$\sigma$ bound on 
the suppression in power from the existing Lyman-$\alpha$ forest data 
from the MIKE/HIRES \cite{Viel:2013apy} and XQ-100 
\cite{2016A&amp;A...594A..91L} datasets used in 
Ref.~\cite{Irsic:2017ixq}. Applied to the MTH, we obtain a bound 
$r_{\text{all}}\lsim 16\,(12)\%$ for $\Delta N_{eff}=0.1\,(0.4)$ that is 
quite insensitive to the $v_B/v_A$ ratio. This agrees with the 
analytical estimate in Eq.~(\ref{eq:rHapprx}) for the average 
suppression at higher $k$-modes. Given the uncertainties involved, we 
should only consider this as a rough guide to the actual 
constraint\footnote{A study of the Lyman-$\alpha$ constraint using the 
SDSS \cite{McDonald:2004eu,McDonald:2004xn} data has been performed in 
Ref.~\cite{Krall:2017xcw} on scenarios in which all dark matter 
particles couple to dark radiation. Although the setup is different, a 
similar study of dark matter scattering off dark radiation may be 
applied to the MTH model.}. The result with $r_{\text{all}}\lesssim 
16\,(12)\%$ is shown as the region above the solid (dashed) blue line in 
Fig.~\ref{fig:LSSbound}.

%%%%%%%%%%%%%%%%%
\section{CMB signals}\label{s.CMB}
%%%%%%%%%%%%%%%%%

The MTH framework predicts a new contribution to the energy density of 
the universe in the form of dark radiation associated with the 
relativistic degrees of freedom in the mirror sector. In order to 
satisfy the current CMB bounds on dark radiation, $\Delta N_{eff} \lsim 
0.45$ \cite{Ade:2015xua,Baumann:2015rya,Brust:2017nmv}, we assume 
asymmetric reheating of the SM bath after the two sectors have 
decoupled. However, while asymmetric reheating increases the energy 
density in the SM sector, it does not erase any pre-existing energy 
density in the mirror sector. Therefore, a small nonvanishing $\Delta 
N_{eff}$ is expected to be a general feature of this framework. This 
dark radiation could potentially be detected at future CMB Stage-IV 
experiments, which are expected to be sensitive to $\Delta N_{eff} 
\gtrsim 0.02$. We treat $\Delta N_{eff}$ as a free parameter, since its 
value depends on the precise details of the asymmetric reheating 
mechanism.

The dark radiation in the MTH is composed of the mirror photon and 
mirror neutrinos. Their relative contributions to $\Delta N_{eff}$ 
depend only on the number of degrees of freedom in the twin bath at the 
time when the mirror neutrinos decouple. For the parameter range of 
interest, $3 \lesssim v_B/v_A \lesssim 5$, the mirror electron goes out 
of the twin bath after the mirror neutrinos have decoupled, just as in 
the visible sector.  Then, even though $\Delta N_{eff}$ itself is a free 
parameter, we have a prediction for the relative contributions of twin 
photons and twin neutrinos to the energy density in dark radiation
 \begin{equation}
\frac{\Delta N^{\hat{\nu}}_{eff}}{\Delta N^{\hat{\gamma}}_{eff}} = \frac{3}{4.4} \;.
 \end{equation} 
 As we now explain, the fact that this ratio is known leads to a 
testable prediction for the CMB. This prediction is a consequence of the 
fact that although twin photons and twin neutrinos both constitute dark 
radiation and contribute to $\Delta N_{eff}$, in detail their effects on 
the CMB are distinct, and can be distinguished. While the twin neutrinos 
free stream, the twin photons are prevented from free streaming by 
scattering off of twin electrons. Scattering and free streaming species 
have different effects on the CMB anisotropies 
\cite{1973ApJ...180....1P,Hu:1995en,Bashinsky:2003tk}. Only after 
recombination has occurred in the mirror sector can twin photons free 
stream. The distinct effects of these two forms of dark radiation on the 
CMB anisotropies can be parametrized in terms of the free streaming 
fraction, defined as
 \begin{equation}
f_\nu = \frac{\rho_{free}}{\rho_r} = \frac{\rho_{free}}{\rho_{free} + \rho_{scatt}} 
 \end{equation}
 Here $\rho_{free}$ represents the energy density in free streaming 
radiation, $\rho_{scatt}$ the energy density in scattering radiation, 
and $\rho_r$ the total energy density in radiation. We use $ 
N^{free}_{eff}$ and $N^{scatt}_{eff}$ to parametrize the energy 
densities in free streaming and scattering radiation in terms of the 
effective number of neutrinos. Then,
 \begin{equation}
f_\nu
= \frac{N^{free}_{eff}}{N^{free}_{eff} + N^{scatt}_{eff}} \;.
 \end{equation} 
 For small $\Delta N_{eff}$, we have \cite{Chacko:2015noa},
 \begin{equation}
 \label{e.fnufSM}
f_\nu - f_\nu^{SM} = \frac{0.41}{3} \left(0.59 \Delta N^{free}_{eff}-0.41 \Delta N^{scatt}_{eff}\right)
 \end{equation}
 For a given $\Delta N_{eff}$, the amplitudes of the CMB modes depend on $f_{\nu}$ as 
\cite{1973ApJ...180....1P,Hu:1995en},
 \begin{equation}
\frac{\delta C_\ell}{C_\ell} = -\frac{8}{15} f_\nu ~.
 \end{equation}
  We see from this that the sign of the correction to the amplitude, relative to the SM, depends on whether the dark radiation free streams or scatters. For a 
given $\Delta N_{eff}$ the locations of the CMB peaks also depend on 
$f_{\nu}$ \cite{Bashinsky:2003tk},
 \begin{equation}
\delta \ell \approx -57 f_\nu \frac{\ell_A}{300} ~.
 \end{equation}
 Here $\ell_A \approx 300$ represents the average angular spacing 
between the CMB peaks at large $\ell$. Once again we see that the sign 
of the shift depends on whether the dark radiation free streams or 
scatters.

The MTH framework predicts the ratio of the energy densities in free 
streaming and scattering species. Prior to recombination in the twin 
sector, we have
 \begin{equation}
\frac{\Delta N^{free}_{eff}}{\Delta N^{scatt}_{eff}} = 
\frac{\Delta N^{\hat{\nu}}_{eff}}{\Delta N^{\hat{\gamma}}_{eff}} =
\frac{3}{4.4},\qquad f_\nu = f_\nu^{SM}.
\label{FSprediction}
 \end{equation}
 Since $\Delta N_{eff}$ and $f_{\nu}$ can be independently determined 
from the CMB, this prediction for $f_{\nu}$ allows the MTH to be 
distinguished from other dark sector scenarios. The point is that while 
a non-zero contribution to $\Delta N_{eff}$ is a characteristic feature 
of any extension of the SM that contains some form of dark radiation, in 
general there is no reason to expect that $f_\nu = f_\nu^{SM}$. The fact 
that in the MTH the free streaming fraction is exactly the same as in 
the SM is because, in this framework, the dark radiation is composed of 
the twin counterparts of the SM photons and neutrinos, with the same 
relative energy densities. The prediction Eq.~(\ref{FSprediction}) 
is not sensitive to the relative fractions of mirror hydrogen and mirror 
helium, but only requires that the energy density in the mirror 
component of dark matter be large enough to sustain acoustic 
oscillations during the CMB epoch, $r_{\rm all} \gtrsim 0.1\%$.

A detailed calculation is needed to obtain a precise prediction for CMB 
observations, since the twin photon becomes free streaming after twin 
recombination. However, the point remains that the MTH makes a unique 
prediction for the ratio of energy densities, $\Delta 
N_{eff}^{free}/\Delta N_{eff}^{scatt}$, prior to twin recombination. 
Since CMB Stage-IV experiments are expected to be very sensitive to not 
just the total energy density in dark radiation, $\Delta N_{eff}$, but 
also the free streaming fraction $f_\nu$ 
\cite{Baumann:2015rya,Brust:2017nmv}, this provides a unique handle to 
test the MTH framework and distinguish it from other possible dark 
sectors.

%%%%%%%%%%%%%%%%%
\section{Conclusions}\label{s.conclusion}
%%%%%%%%%%%%%%%%%

In this paper, we have explored the cosmological signals associated with 
the mirror baryons, electrons, photons, and neutrinos in the MTH model. 
We have worked in a framework in which the cosmological problems of the 
original MTH proposal are assumed to be solved by late time asymmetric 
reheating after the two sectors have decoupled. We have primarily 
focused on the case in which the discrete $Z_2$ symmetry that relates 
the two sectors is only softly broken. Then, in order for the little 
hierarchy problem to be addressed, the masses of particles in the mirror 
sector are restricted to lie in a limited range. This means that, 
although many of the late time thermal processes, such as BBN and 
recombination, are sensitive to the particle masses and couplings, we 
have still been able to draw a clear picture of the cosmology of the MTH 
framework, which exhibits several characteristic features. Therefore 
cosmological observations may allow this class of models, which are 
difficult to test in collider experiments, to be discovered.

From a study of TBBN, we have found that, in contrast to the SM, 
$\approx 75\%$ of the mass density in mirror particles density is 
contained in helium. Mirror hydrogen and helium remain ionized until 
close to the time of matter-radiation equality and, as in the SM, 
scattering off twin electrons and photons generates TBAO that suppress 
matter density perturbations. Although mirror helium recombines into 
neutral atoms before mirror hydrogen, its relatively larger abundance 
means that its impact on LSS, especially on shorter wavelengths, cannot 
be neglected. TBAO is ended by recombination in the mirror sector, 
leading to an oscillatory feature in the matter power spectrum in 
$k$-space with a frequency lower than that of SM BAO. This may allow an 
observation of TBAO in the near future in LSS observations. Current 
observations allow up to $ 5\,(10)\%$ of the matter density to come from 
twin hydrogen and helium for $\Delta N_{eff}=0.4\,(0.1)$. In the absence 
of a signal, future LSS measurements will be able to tighten this bound 
to the $ 1\%$ level.

In the MTH framework, we expect observable contributions to $\Delta 
N_{eff}$ from mirror photons and neutrinos. Although $\Delta N_{eff}$ 
itself is a free parameter, the relative energy densities in these two 
species are known. The mirror photons and mirror neutrinos have distinct 
effects on the CMB because, until recombination, the twin photons 
scatter off the twin electrons, which prevents them from free streaming. 
For any given $\Delta N_{eff}$, this leads to a prediction for the 
corrections to the heights and locations of the CMB peaks that can be 
potentially be tested in future experiments.

Our analysis has primarily focused on the case in which the discrete 
$Z_2$ symmetry is only softly broken, so that the relative abundances of 
twin hydrogen and helium are predicted. However, for the purposes of 
comparison we have also explored scenarios in which the nuclei in the 
mirror sector are composed entirely of hydrogen, or entirely of helium. 
In particular, our studies capture the features of the important case in 
which, because of hard breaking of the discrete $Z_2$ symmetry in the 
Yukawa couplings of the light quarks, the mirror neutron is lighter than 
the mirror proton, and constitutes the dominant component of the 
observed dark matter, while mirror helium represents an acoustic 
subcomponent that gives rise to the signals we consider. Interestingly, 
we find that because hydrogen and helium recombine at different times, 
the matter power spectrum of the framework in which both nuclei are 
present exhibits distinctive features that may allow it to be 
distinguished from the case of atomic dark matter with just a single 
type of nucleus.

 While we have focused on cosmological signals of the MTH framework, it 
is worth keeping in mind that this scenario can also give rise to 
striking astrophysical signals. During the formation of the galactic 
halo, the mirror baryonic component of dark matter in the Milky Way may 
have become re-ionized. If the ionized mirror baryons subsequently 
dissipated enough energy, they may have collapsed into a dark disk, as 
in the scenario discussed in \cite{Fan:2013yva}. The details of this 
process depend sensitively on the MTH parameters, and on the initial 
distribution of dark matter in the galaxy. If the mirror baryons do form 
a second disk aligned with our own, current Gaia 
observations~\cite{Schutz:2017tfp} already constrain the mirror baryons 
to constitute less than $\sim 1\%$ of the dark matter density. However, 
a careful study is needed to draw robust conclusions about the allowed 
parameter space. The distribution of mirror baryons and electrons within 
our galaxy will also determine how they might be discovered at current 
or future dark matter direct detection experiments. In the MTH 
framework, the SM and mirror sectors interact through the Higgs portal. 
Unfortunately, the resulting signal is far too small for direct 
detection of mirror hydrogen or mirror helium nuclei in any current or 
proposed experiment. However, in the event of a small kinetic mixing 
between the SM photon and its twin counterpart, the mirror baryons and 
electrons will acquire a tiny electric charge. The mirror sector can 
then interact with the SM through this portal, opening a new pathway for 
direct detection. Avoiding recoupling of the mirror sector after 
asymmetric reheating via $e\hat{e}$ scattering for $T \gtrsim \mev$ 
leads to an upper bound on the kinetic mixing parameter, $\epsilon 
\lesssim 10^{-9}$. In the MTH model, no kinetic mixing is generated 
through 3-loop order~\cite{Chacko:2005pe}, and therefore even such small 
values of $\epsilon$ are radiatively stable. It follows that this bound 
can naturally be satisfied provided that the contributions to $\epsilon$ 
from UV physics are sufficiently small. Therefore $\epsilon$ of order 
$10^{-9}$, corresponding to \emph{nano-charged dark matter}, constitutes 
a natural sensitivity goal for future direct detection experiments. % 
Such small kinetic mixings are also compatible with supernova 
bounds~\cite{Chang:2018rso}. % While some aspects of nano-charged dark 
matter direct detection have been previously studied in the context of 
mirror models (see \cite{Foot:2014mia} and references contained 
therein), the signals again depend sensitively on the details of the 
hidden sector. These striking astrophysical signals of the MTH 
framework, which clearly warrant further study, will be the subject of a 
companion paper~\cite{MTHddinprogress}.

The MTH scenario with minimal $Z_2$ breaking avoids almost all collider 
bounds, since it addresses the hierarchy problem with top partners that 
are neutral under all the SM forces. Our study demonstrates that in this 
framework the mirror baryons, electrons, photons and neutrinos can give 
rise to a rich plethora of distinctive cosmological and astrophysical 
signatures. Taken together, these may allow the nature of the hidden 
sector, and its possible relation to the SM and the hierarchy problem, 
to be discovered and probed in considerable detail. This opens the door 
to the tantalizing possibility that the first hints of naturalness may 
come from the sky, rather than from colliders.

%%%%%%%%%%%%
 \begin{acknowledgments}

 We thank 
 Masha Baryakhtar,
 Paulo Bedaque,
 Thomas Cohen, 
 Rouven Essig, 
 Junwu Huang,
 Robert Lasenby,
 Shmuel Nussinov,
 Tien-Tien Yu 
 and Yue Zhao
for helpful discussions. 
 ZC is supported in part by the National Science Foundation under Grant 
Number PHY-1620074, the Fermilab Intensity Frontier Fellowship and the 
Visiting Scholars Award \#17-S-02 from the Universities Research 
Association. The work of DC, MG and YT was supported in part by the 
National Science Foundation under grant PHY-1620074, and by the Maryland 
Center for Fundamental Physics. ZC would like to thank the Fermilab 
Theory Group for hospitality during the completion of this work. YT 
thanks the Aspen Center for Physics, which is supported by National 
Science Foundation grant PHY-1066293.
 \end{acknowledgments}

%%%%%%%%%%%%%%%%%%%%%%%%%%%%%%%%%%%%%%%

%%%%%%%%%%%%%%%%%%%%%%%%%%%%%%%%%%%%%%%
\bibliography{MTH_cosmo}
\bibliographystyle{JHEP}

%%%%%%%%%%%%%%%%%%%%%%%%%%%%%%%%%%%%%%%

\end{document}